\documentclass[]{mn2e}

\usepackage{epsfig}

\title[The origin of sdB stars (II)] {The Origin of Subdwarf B Stars (II)}

\author[Han et al.]
{Z.~Han$^1$\thanks{E-mail: zhanwen@public.km.yn.cn}, Ph.~Podsiadlowski$^2$,
P. F. L. Maxted$^3$, T. R. Marsh$^4$\\
\it
$^1$ National Astronomical Observatories / Yunnan Observatory, 
the Chinese Academy of Sciences, P.O.Box 110, Kunming, 650011, China\\
$^2$ University of Oxford, Department of Astrophysics, Keble Road, Oxford,
OX1 3RH\\
$^3$ School of Chemistry and Physics, Keele University, Staffordshire, ST5 5BG\\
$^4$ University of Southampton, Department of Physics {\rm\&} Astronomy,
Highfield, Southampton, S017 1BJ
}

\begin{document}
\maketitle

\begin{abstract}
We have carried out a detailed binary populations synthesis (BPS)
study of the formation of subdwarf B (sdB) stars and related objects
(sdO, sdOB stars) using the latest version of the BPS code developed
by Han et al.\ (1994, 1995a, 1995b, 1998, 2001).  We systematically
investigate the importance of the five main evolutionary channels in
which the sdB stars form after one or two common-envelope (CE) phases,
one or two phases of stable Roche-lobe overflow (RLOF) or as the
result of the merger of two helium white dwarfs (WD) (see Han et al.\
2002, Paper I).  Our best BPS model can satisfactorily explain the
main observational characteristics of sdB stars, in particular their
distributions in the orbital period -- minimum companion mass ($\log
P$ - $M_{\rm comp}$) diagram and in the effective temperature --
surface gravity ($T_{\rm eff}$ - $\log g$) diagram, their
distributions of orbital period, $\log (g\theta^4)$
($\theta=5040\,{\rm K}/T_{\rm eff}$) and mass function, their binary
fraction and the fraction of sdB binaries with WD companions, their
birthrates and their space density.  We obtain a Galactic
formation rate for sdB stars of 0.014\,--\,$0.063\,{\rm yr}^{-1}$ with
a best estimate of $\sim 0.05\, {\rm yr}^{-1}$ and a total number in
the Galaxy of 2.4\,--\,$9.5 \times 10^6$ with a best estimate of $\sim
6\times 10^6$; half of these may be missing in observational surveys
due to selection effects. The intrinsic binary fraction is 76 to 89
percent, although the observed frequency may be substantially lower
due to the selection effects. The first CE ejection channel, the
first stable RLOF channel and the merger channel are intrinsically the
most important channels, although observational selection effects tend
to increase the relative importance of the second CE ejection and 
merger channels. We also predict a distribution of masses for
sdB stars that is wider than is commonly assumed and that some sdB
stars have companions of spectral type as early as B.  The percentage
of A type stars with sdB companions can in principle be used to
constrain some of the important parameters in the binary evolution
model.  We conclude that (a) the first RLOF phase needs to be more
stable than is commonly assumed, either because the critical mass
ratio $q_{\rm crit}$ for dynamical mass transfer is higher or because
of tidally enhanced stellar wind mass loss; (b) mass transfer in the
first stable RLOF phase is non-conservative, and the mass lost from
the system takes away a specific angular momentum similar to that of
the system; (c) common-envelope ejection is very efficient.


\end{abstract}


\section{Introduction}

Hot subdwarfs are defined as stars that are located below the upper
main sequence in the Hertzsprung-Russell diagram (HRD); this class of
objects includes subdwarf B (sdB), subdwarf O (sdO) and subdwarf OB
(sdOB) stars \cite{vau87,kil88}. The majority of hot subdwarfs in
photographic surveys are sdB stars, which we use as a collective term
for all hot subdwarfs (i.e.\  including sdO and sdOB stars).

Due to their ubiquity, sdB stars play an important role in the study of
the Galaxy \cite{gre86}. Pulsating sdB stars can be
used as standard candles and hence distance indicators \cite{kil99}.
In external galaxies, they may provide the dominant source of
ultraviolet (UV) radiation in old stellar populations, such as giant
elliptical galaxies. The UV excess, or ``UV upturn'', in old
populations has be used as an age indicator of giant elliptical
galaxies using an evolutionary population synthesis approach
\cite{bro97,yi97,yi99}, which has important cosmological applications.
Despite of their importance, their origin has still remained somewhat
of a puzzle, and they provide an important link in our understanding
of both single and binary stellar evolution theory.

sdB stars are generally considered to be core helium-burning stars
with extremely thin hydrogen envelopes ($<0.02\,M_\odot$) and masses
around $0.5\,M_\odot$ \cite{heb86,saf94}, as has recently been
confirmed asteroseismologically in the case of PG 0014+067 (Brassard
et al.\ 2001). Maxted et al.\ (2001) showed that more than half of the
sdB stars in their selected sample are members of close binaries.
There have been many theoretical investigations on the formation of
sdB stars in the past. Webbink \shortcite{web84} and Iben \& Tutukov
\shortcite{ibe86} proposed that the coalescence of two helium white
dwarfs (WDs) may produce sdB stars. Tutukov \& Yungelson (1990)
estimated that this would be the dominant formation channel. D'Cruz et
al.\ \shortcite{dcr96} argued that an enhanced stellar wind near the
tip of the first giant branch (FGB) can result in the formation of sdB
stars, while Sweigart \shortcite{swe97} suggested that helium mixing
driven by internal rotation may account for such enhanced mass loss.
All of these channels produce sdB stars that are either single or in
wide, non-interacting binaries.  sdB stars in binaries can form
through various binary channels, involving either stable and
conservative mass transfer (Mengel, Norris \& Gross 1976) or dynamical
mass transfer and common-envelope evolution (Paczy\'nski 1976).

To understand the formation of sdB stars, Han et al.\
\shortcite{han02} (hereafter Paper I) have performed a systematic
study of the various binary evolution channels that can produce sdB
stars. Using simplified binary population synthesis (BPS) simulations
for some of the channels, they showed that all of these proposed
channels proposed are viable in principle. The purpose of the present
paper is to quantitatively assess the relative importance of the
various channels by performing a full binary population synthesis
study and by constraining the theoretical models from the observed
properties of the population of sdB stars.

The outline of the paper is as follows.  In sections 2 and 3 we
briefly summarize the observations of sdB stars and the principal
formation channels, respectively.  We describe our BPS code and the
model parameters in section 4 and constrain some of the main model
parameters in section 5.  In section 6 we carry out a large number of
BPS simulations and present the main results, which are then discussed
in detail in section 7.  The conclusions in section 8 summarize the main
findings of the present study.

\section{Observations of sdB stars}

There have been extensive observations of sdB stars during the past
decades. Magnitude-limited samples of sdB stars, selected by colour,
have been made from the Palomar Green (PG) survey \cite{gre86} ($B\sim
16.1$) and the Kitt Peak Downes (KPD) survey \cite{dow86} ($B\sim
15.3$). Saffer et al.\ (1994) measured atmospheric parameters,
such as effective temperature, surface gravity and photospheric helium
abundance, for a sample of 68 sdB stars.  Ferguson, Green \& Liebert
\shortcite{fer84} found 19 sdB binaries with main-sequence (MS)
companions from the PG survey and derived a binary frequency of about
50 percent. Allard et al.  \shortcite{all94} found 31 sdB binaries
from 100 candidates chosen from the PG survey and the KPD
colourimetric survey, and estimated that 54 to 66 percent of sdB stars
are in binaries with MS companions after taking selection effects into
account.  Thejll, Ulla \& MacDonald \shortcite{the95} and Ulla \&
Thejll \shortcite{ull98} found that more than half of their sdB star
candidates showed infrared flux excesses, indicating the presence of
binary companions.  Aznar Cuadrado \& Jeffery \shortcite{azn01}
obtained atmospheric parameters for 34 sdB stars from spectral energy
distributions and concluded that 15 of these were single and 19
binaries with MS companions. All of these observations indicated that
more than half of sdB stars were in binaries. (Note, however, that
some of the MS `companions' to sdB stars are optical doubles and are not
physically related.)

More recently, it has become possible to determine some of the orbital
parameters, such as orbital periods and mass functions, for a significant
sample of close sdB binaries
\cite{jef98,koe98,saf98,kil99,mor99,oro99,woo99,max00a,max00b,max01,nap01,dre02,heb02,mor02a,mor02b}. 
In particular, Maxted et al.\ \shortcite{max01} concluded that more
than two thirds of their candidates were binaries with short orbital
periods from hours to days, and that 7 of 11 sdB binaries with known
companion types had WD companions. Since this study has very
well-defined selection criteria, it provides an excellent data set to
help constrain the theoretical models. The main selection effects in
the data set are: a) a selection in the PG survey against sdB stars
with companions of spectral type G and K (which show composite
spectra) and companions of earlier spectral types (which dominate the
optical light output); b) the major fraction of candidates was
selected from a narrow strip in the $T_{\rm eff}-\log g$ diagram for
sdB stars with masses of $\sim 0.5\,M_\odot$ which are believed to be
in the core helium-burning phase; c) the radial-velocity
semi-amplitudes ($K$) of all sdB binaries with known orbital periods
are larger than 30 km\,s$^{-1}$. We therefore exclude in some of our
comparisons all systems with smaller semi-amplitudes.  This selects
the sample against systems with long orbital periods and/or low
companion masses.  In principle, orbital periods for binaries with
semi-amplitudes as low as 10\,km\,s$^{-1}$ can be detected, but
because of their typically long expected orbital periods no periods
have yet been determined observationally
\cite{max01}\footnote{From a rigorous statistical 
point of view, it would be more correct to introduce a separate
period-selection criterion. However, since this is not entirely
straightforward we chose this simpler criterion and note that this
criterion was not actually used to constrain any of the theoretical
parameters in this paper.}.  We shall refer to these selection effects
as the GK selection effect (a), the strip selection effect (b) and the
K selection effect (c), respectively.

\section{Binary formation channels}

We consider sdB stars to be core helium-burning stars with masses
around $0.5\,M_\odot$ with extremely thin hydrogen-rich envelopes 
\cite{heb86,saf94}. The main binary channels that can produce
sdB stars were discussed in detail in Paper I. Here we restrict
ourselves to summarizing some of their main features.

\subsection{The first CE ejection channel}

In this channel, the primary component, i.e. the initially more
massive star of the binary, experiences dynamical mass transfer on the
FGB.  This leads to a CE and a spiral-in phase, typically leaving a
very close binary after the envelope has been ejected. If the core of
the giant still ignites helium it produces a sdB star in a
short-period binary with a main-sequence companion.

Depending on the initial mass of the primary, one has to distinguish
between two sub-channels. If the initial mass is below the helium
flash mass $M_0$, i.e. the maximum ZAMS mass below which a star
experiences a helium flash at the tip of the FGB ($M_0\sim
1.99\,M_\odot$ for Pop I, $M_0\sim 1.80\,M_\odot$ for $Z=0.004$, see
\S~3.2 of Paper I), the primary must fill its Roche lobe when it is
already quite close to the tip of the FGB in order to be able to
ignite helium. All sdB stars formed through this channel should have
masses just below the critical core mass for the helium flash and have
a mass distribution peaked around $0.46\,M_\odot$. The orbital
period distribution typically ranges from 0.05 to $\ga 40\,$d .
 
If the ZAMS mass is higher than the helium flash mass, the primary
does not have to be close to the tip of the FGB since more massive
primaries will ignite helium (in this case under non-degenerate
conditions) even if they lose their envelopes when passing through the
Hertzsprung gap. However, since the envelopes of stars in the
Hertzsprung gap are much more tightly bound than on the FGB, systems
that experience dynamical mass transfer in the Hertzsprung gap are
more likely to merge completely than to survive as short-period
binaries. As a consequence this sub-channel does not contribute much
to the formation of sdB stars, although it should be noted that these
would generally contain sdB stars of lower mass (as low as $\sim
0.33\,M_\odot$) and tend to have very short orbital periods.

\subsection{The first stable RLOF channel}

If the first mass-transfer phase is stable, the primary will also lose
most of its envelope producing a sdB star with a MS companion but in
this case in a wide orbit with orbital periods between $\sim 0.5$ and
2000\,d. The orbital period depends on how angular momentum is lost
from the system with the shortest periods resulting from systems that
experience stable RLOF near the beginning of the Hertzsprung gap.

Similarly to the previous channel, one has to distinguish between
two sub-channels.  If the primary has a ZAMS mass below the helium 
flash mass, Roche-lobe overflow has to occur near the tip of the 
FGB which again leads to a sharp peak in the mass distribution around 
$0.46\,M_\odot$.

If the primary has a ZAMS mass larger than the helium flash mass and
the system experiences stable RLOF in the Hertzsprung gap (so-called
early case B mass transfer), this also leads to the formation of a sdB
star, as was shown in detailed binary evolution calculations by Han,
Tout \& Eggleton \shortcite{han00}.  We adopt these models to define
the evolution for this sub-channel.  In this case, the mass of the sdB
star can have a very wide range from $0.33\,M_\odot$ to $1.1\,M_\odot$,
although the more massive sdB stars are less likely because of their
lower realization probability due to the initial mass function.

\subsection{The second CE ejection channel}

This channel is similar to the first CE ejection channel, except that
the companion to the giant is already a white dwarf.  This can lead to
a shorter orbital period of the sdB binary after the CE ejection since
the WD companion has a much smaller radius than a MS star and a WD can
penetrate much deeper into the CE and cause its ejection; i.e. it can
avoid the complete merging of the two components. Therefore sdB stars
from this channel have a wider range of orbital periods, and their
companions are WDs.

Again there are two sub-channels depending on the initial mass of the
giant. However, unlike the first CE ejection channel, the more massive
channel contributes more to the sdB population since it is easier to
eject the envelope of a star in the Hertzsprung gap if the companion
is a white dwarf. The masses of sdB stars from the first sub-channel
are $\sim 0.46\,M_\odot$, while the masses of those from the second are
$\sim 0.35\,M_\odot$.

\subsection{The second stable RLOF channel}

This channel is similar to the first stable RLOF channel.  However, in
order to have stable RLOF, the ZAMS mass of the giant is very
restricted (the mass ratio of the giant to the WD, $M_{\rm RG}/M_{\rm
WD}$, has to be below a value of $\sim 1.1$\,--\,1.3; see Table~3 of
Paper I). This generally requires very massive WD companions. Since
these are very rare, this channel is unlikely to contribute much to
the sdB star population.  In fact in our simulations, we do not
produce any sdB stars from this channel since the WD companions tend
not to be sufficiently massive\footnote{However, this would be
different if we had included tidally enhanced wind mass loss, since
this would reduce the minimum mass of the white dwarf for dynamically
stable mass transfer.}.

\subsection{The helium WD merger channel}

Binaries containing two helium WDs may be produced after
either two CE phases or one stable RLOF phase and one CE phase 
\cite{web84,ibe86,han98}.
If their orbital period is sufficiently short, the systems will shrink
due to gravitational wave radiation, and the two helium white dwarfs
may coalesce. If the merger product ignites helium, this again leads
to the formation of a single sdB star (Saio \& Jeffery 2001) with a
fairly wide mass distribution ($\sim 0.4$ to $0.6\,M_\odot$; see Paper
I).

\section{The Binary population synthesis code}

\subsection{Code description}
   
The BPS code used here was originally developed in 1994 and has been
updated regularly ever since
\cite{han94,han95a,han95c,han95b,han98,han01}. The main input into the
code is a grid of stellar models. We use 3 grids of older models for
metallicities $Z=0.02$, 0.004 and 0.001, which do not include
convective overshooting or stellar winds. For the purpose of the
present study, we calculated 6 new grids for $Z=0.02$ and 0.004. These
are smaller and cover a smaller range of masses -- as 
appropriate for the study of sdB stars. The new grids include stellar
winds and convective overshooting (see Paper I for a more detailed
description).

The code needs to model the evolution of binary stars as well as of
single stars.  Single stars are evolved according to the model grids,
while the evolution of binaries is more complicated due to the
occurrence of RLOF. A binary usually experiences two phases of RLOF;
the first when the primary fills its Roche lobe which may produce a WD
binary and the second when the secondary fills its Roche lobe.

The mass gainer in the first RLOF phase is most likely a MS star. If
the mass ratio $q=M_1/M_2$ at the onset of RLOF is lower than a
critical value $q_{\rm crit}$, RLOF is stable
\cite{pac65,pac69,pla73,hje87,web88,sob97,han01}. For systems 
experiencing their first phase of RLOF in the Hertzsprung gap, we use
$q_{\rm crit}=3.2$ as is supported by a simple model due to
P.P. Eggleton (private communication) and by detailed binary evolution
calculations Han et al.\ (2000).  For the first RLOF phase on the FGB or AGB
we use three different prescriptions to examine the consequences of
varying this important criterion:
\begin{enumerate}
\item
$q_{\rm crit}=0.362+ {1/[{3(1-m_{\rm c})}]}$, 
where $m_{\rm c}$ is the core mass fraction. This criterion was
derived by Hjellming \& Webbink \shortcite{hje87} and 
Webbink \shortcite{web88} for conservative mass transfer 
and a mass donor modelled as a polytrope (also see Soberman et al.\ 1997).
For examples involving non-conservative mass transfer, see Han et al.\ 
(2001). 
\item
$q_{\rm crit}=1.2$ 
\item
$q_{\rm crit}=1.5$
\end{enumerate}
We assume that a fraction $\alpha_{\rm RLOF}$ of the mass lost from
the primary is transferred onto the gainer, while the rest is lost
from the system ($\alpha_{\rm RLOF}=1$ means that RLOF is
conservative).  Note that we assume that mass transfer during the
main-sequence phase is assumed to be always conservative.  The mass
lost from the system also takes away angular momentum, for which we
adopt two different choices:
\begin{enumerate}
\item[(i')]
the mass lost takes away the same specific angular momentum as
the orbital angular momentum of the primary
\item[(ii')] 
the mass lost takes away a specific angular momentum $\alpha$ in units
of the specific angular momentum of the system. The unit is expressed
as $2\pi a^2/P$, where $a$ is the separation and $P$ is the orbital
period of the binary (see Podsiadlowski, Joss \& Hsu 1992, hereafter
PJH, for details).
\end{enumerate}  

Stable RLOF usually results in a wide WD binary. Some of the wide WD
binaries may contain sdB stars if RLOF occurs near the tip of the
FGB. RLOF near the tip of the FGB is not likely to be stable if one
uses the polytropic criterion (criterion i) since $q_{\rm crit}$ is
generally less than 1 and since we do not explicitly include tidally
enhanced stellar winds \cite{tou88,egg89b,han95b}. When using a larger
value for $q_{\rm crit}$, the number of systems experiencing stable
RLOF increases significantly. To some degree this is equivalent to
including a tidally enhanced stellar wind.  Moreover, the full binary
calculations presented in Paper I demonstrate that a larger value of
$q_{\rm crit}$ is the more appropriate one to use. These calculations 
gave a typical $q_{\rm crit}\sim 1.2$ (see Table 3 of Paper I), very 
different from what the polytropic model predicts.

If RLOF is dynamically unstable, a CE may be formed \cite{pac76}, and
if the orbital energy deposited in the envelope can overcome its
binding energy, the CE may be ejected. For the CE ejection criterion,
we introduced two model parameters, $\alpha_{\rm CE}$ for the common
envelope ejection efficiency and $\alpha_{\rm th}$ for the thermal
contribution to the binding energy of the envelope, which we write as
\begin{equation}
\alpha_{\rm CE}\,|\Delta E_{\rm orb}| > |E_{\rm gr} + \alpha_{\rm th}
\,E_{\rm th}|,
\end{equation}
where $\Delta E_{\rm orb}$ is the orbital energy that is released,
$E_{\rm gr}$ is the gravitational binding energy and $E_{\rm th}$ is
the thermal energy of the envelope.  Both $E_{\rm gr}$ and $E_{\rm
th}$ are obtained from full stellar structure calculations (see Han,
Podsiadlowski \& Eggleton 1994, hereafter HPE, for details; also see
Dewi \& Tauris 2000) instead of analytical approximations.  CE
ejection leads to the formation of a close WD binary and may give rise to
the formation of a sdB star in a short-period system with a MS companion.

The WD binary formed from the first RLOF phase continues to evolve,
and the secondary may fill its Roche lobe as a red giant. The system
then experiences a second RLOF phase.  If the mass ratio at the onset
of RLOF is greater than the critical value $q_{\rm crit}$ given in
Table 3 of Paper I, RLOF is dynamically unstable, leading again to a
CE phase.  If the CE is ejected, a sdB star may be formed (see
\S~3.3). The sdB binary has a short orbital period and a WD companion.
However, RLOF may be stable if the mass ratio is sufficiently small.
In this case, we assume that the mass lost from the mass donor is all
lost from the system, carrying away the same specific angular momentum
as pertains to the WD companion. Stable RLOF may then result in the
formation of a sdB binary with a WD companion and a long orbital
period (typically $\sim 1000\,{\rm d}$).

If the second RLOF phase results in a CE phase and the CE is ejected,
a double white dwarf system is formed \cite{web84,ibe86,han98}.  Some
of the double WD systems contain two helium WDs.  Angular momentum
loss due to gravitational radiation may then cause the shrinking of
the orbital separation until the less massive white dwarf starts to
fill its Roche lobe. This will lead to its dynamical disruption if
\begin{equation}
q \ga 0.7 - 0.1 (M_2/\,M_\odot) 
\end{equation}
or $M_1\ga 0.3\,M_\odot$, where $M_1$ is the mass of the donor
(i.e. the less massive WD) and $M_2$ is the mass of the gainer
\cite{han99}. This is expected to always lead to a complete merger of
the two white dwarfs. The merger can also produce a sdB star, but in
this case the sdB star is a single object. If the lighter WD is not
disrupted, RLOF is stable and an AM CVn system is formed.

In this paper, we do not include a tidally enhanced stellar wind
explicitly as was done in Han et al.\ (1995) and Han (1998).  Instead
we use a standard Reimers wind formula (Reimers 1975) with $\eta =
1/4$ \cite{ren81,ibe83,car96} which is included in our new stellar
models. This is to keep the simulations as simple as possible,
although the effects of a tidally enhanced wind can to some degree be
implicitly included by using a larger value of $q_{\rm crit}$.  We
also employ a standard magnetic braking law \cite{ver81,rap83} where
appropriate (see Podsiadlowski, Han \& Rappaport [2002] for details and
further discussion).

\subsection{Monte Carlo simulation parameters}

To estimate the importance of each evolutionary channel for the production
of sdB stars, we have performed a series of Monte Carlo simulations where
we follow the evolution of a sample of a million binaries 
according to our grids of stellar models. In addition, the simulations
require as input the star formation rate (SFR), the initial mass function
(IMF) of the primary, the initial mass-ratio distribution and
the distribution of initial orbital separations.

(1) The SFR is taken to be constant over the last 15\,Gyr.

(2) A simple approximation to the IMF of Miller \& Scalo \shortcite{mil79} 
 is used;
the primary mass is generated with the formula of Eggleton, Fitchett
\& Tout \shortcite{egg89a},
\begin{equation}
M_1={0.19X\over (1-X)^{0.75}+0.032(1-X)^{0.25}}, 
\end{equation}
where $X$ is a random number uniformly distributed between 0 and 1.
The adopted ranges of primary masses are 0.8 to $100.0\,M_\odot$. The
studies by Kroupa, Tout
\& Gilmore \shortcite{kro93} and Zoccali et al.\ \shortcite{zoc00}
 support this IMF.

(3) The mass-ratio distribution is quite controversial. We mainly
take a constant mass-ratio distribution \cite{maz92,gol94},
\begin{equation}
n(1/q)=1,\qquad  0\leq 1/q \leq 1, 
\end{equation}
where $q=M_1/M_2$. As an alternative mass-ratio distribution we also
consider the case where both binary components are chosen randomly and
independently from the same IMF.

(4) We assume that all stars are members of binary systems and that the
distribution of separations is constant in $\log a$
($a$ is the separation) for wide
binaries and falls off smoothly at close separations:
\begin{equation}
an(a)=\cases {\alpha_{\rm sep}({a\over a_0})^m, & $a\leq a_0$;\cr
               \alpha_{\rm sep}, & $a_0 < a < a_1$\cr } 
\end{equation}
where $\alpha_{\rm sep} \approx 0.070$, $a_0=10\,R_\odot$,
$a_1=5.75\times 10^6\,R_\odot=0.13\,{\rm pc}$, and $m\approx 1.2$.
This distribution implies that there is an equal number of wide binary
systems per logarithmic interval and that approximately 50 per cent
of stellar systems are binary systems with orbital periods less than
100\,yr.

\section{Observational constraints}

\begin{figure}
\epsfig{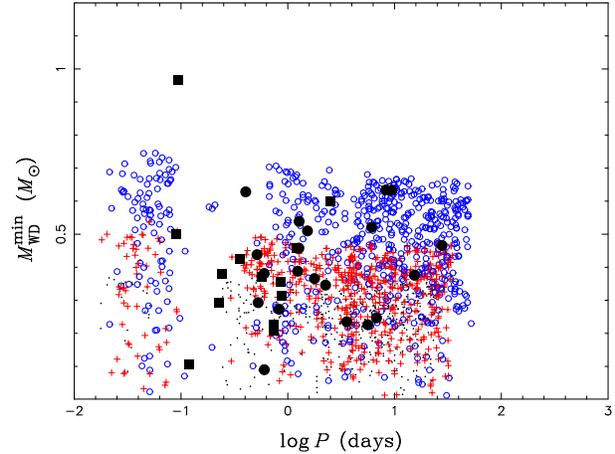}
\caption{
Minimum white-dwarf mass, $M_{\rm WD}^{\rm min}$, versus orbital period, $P$,
for sdB stars in short-period binaries (filled symbols; Maxted et al.\ 2001;
Morales-Rueda et al.\ 2002a,b) 
and the simulated distribution of sdB stars produced from the 2nd 
CE ejection channel. Filled squares indicate observed sdB stars
with known WD companions, filled circles sdB binaries where the nature
of the companion is unknown. The symbols for the simulated systems
indicate the actual masses of the white dwarfs 
(dots: $0.25\leq M_{\rm WD}\leq 0.35\,M_\odot$,
 pluses: $0.35< M_{\rm WD}\leq 0.45\,M_\odot$,
 circles: $0.55\leq M_{\rm WD}\leq 0.65\,M_\odot$). 
The simulation shown uses a standard set of BPS assumptions and 
CE parameters $\alpha_{\rm CE}=0.75$ and $\alpha_{\rm th}=0.75$ 
(i.e. similar to our best model parameters; see section~7.4).
}
\label{p-mwd}
\end{figure}

\begin{figure}
\epsfig{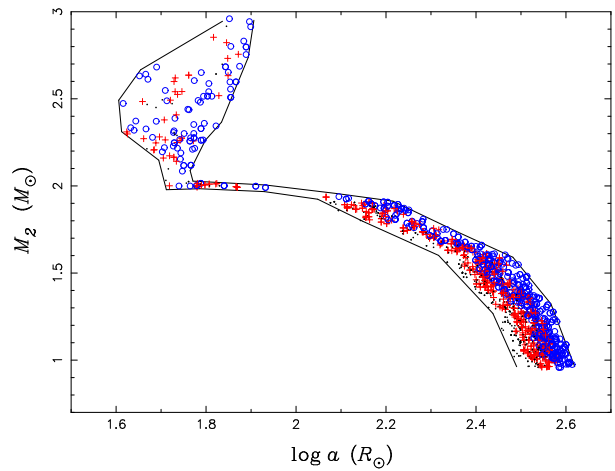}
\caption{The distribution of the progenitors of sdB binaries
(before the 2nd CE phase) in the $a$--$M_2$ plane that produces
the distribution in Figure~\ref{p-mwd}, where $a$ is the
orbital separation and $M_2$ is the initial mass of the sdB star on the 
main sequence. 
The symbols indicate the mass of the white dwarfs (as in Fig.~\ref{p-mwd}).
Solid curves mark the boundaries that produces sdB stars.
}
\label{wdbinary}
\end{figure}

\begin{figure}
\epsfig{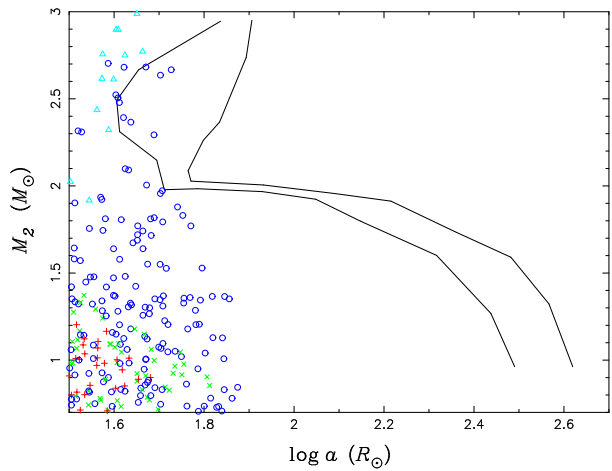}
\caption{
The distribution of Pop I WD binaries in the $a$--$M_2$ plane after CE
ejection where the first mass-transfer phase is dynamically unstable
(with $\alpha_{\rm CE}=0.75$ and $\alpha_{\rm th}=0.75$).  The symbols
indicate the masses of the white dwarfs (dots: $0.25\leq M_{\rm
WD}\leq 0.35\,M_\odot$, pluses: $0.35< M_{\rm WD}\leq 0.45\,M_\odot$,
crosses: $0.45< M_{\rm WD}\leq 0.55\,M_\odot$, circles: $0.55< M_{\rm
WD}\leq 0.65\,M_\odot$, triangles: $M_{\rm WD}> 0.65\,M_\odot$).  }
\label{not-ce}
\end{figure}

\begin{figure}
\epsfig{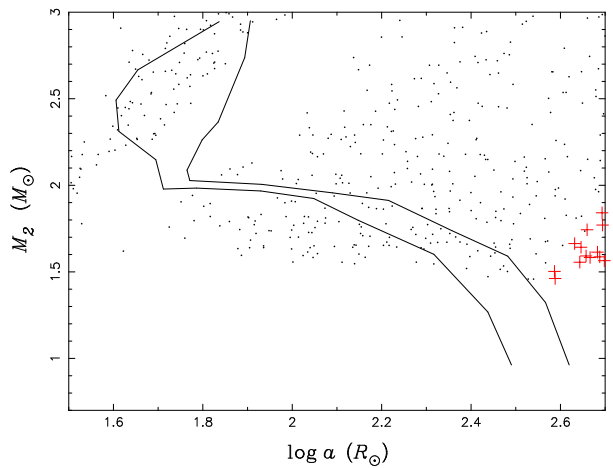}
\caption{The distribution of Pop I WD binaries in the $a$--$M_2$
plane, similar to Figure~\ref{not-ce}, but where the
first mass-transfer is stable leading to conservative RLOF.
}
\label{non-cons}
\end{figure}

\begin{figure}
\epsfig{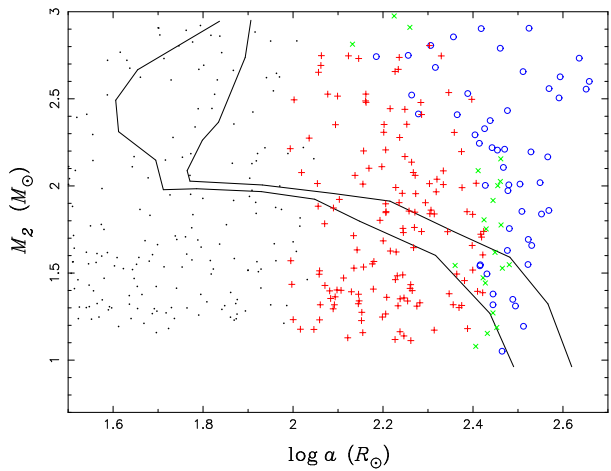}
\caption{Similar to Figure~\ref{not-ce}, where the first mass-transfer
phase is stable but non-conservative with $\alpha_{\rm RLOF}=0.5$. The
mass lost from the system is assumed to take away the same specific
angular momentum as pertains to the system (i.e. $\alpha=1.0$ in PJH's
formalism).
}

\label{a-rlof}
\end{figure}

\begin{figure}
\epsfig{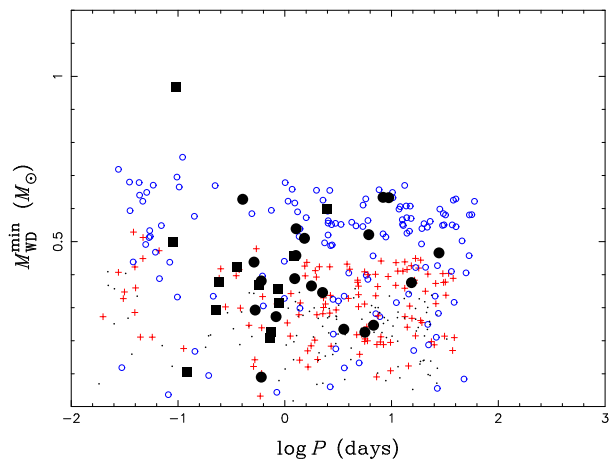}
\caption{
Similar to Figure~\ref{p-mwd}, but for
sdB stars with $Z=0.004$.
}
\label{pmwdz004}
\end{figure}

\begin{figure}
\epsfig{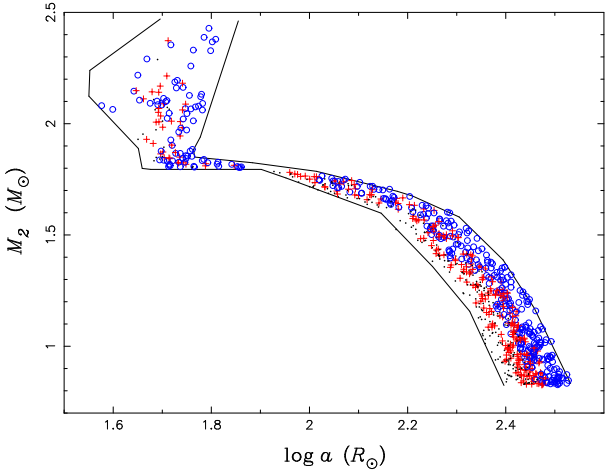}
\caption{
The distribution of the progenitors of sdB binaries
(before the 2nd CE phase) in the $a$--$M_2$ plane that produces
the distribution in Figure~\ref{pmwdz004} (similar to Fig.~\ref{wdbinary}).
The symbols indicate the mass of the white dwarfs 
(dots: $0.25\leq M_{\rm WD}\leq 0.35\,M_\odot$,
 pluses: $0.35< M_{\rm WD}\leq 0.45\,M_\odot$,
 circles: $0.55\leq M_{\rm WD}\leq 0.65\,M_\odot$). 
Solid curves mark the boundaries that produces sdB stars.
}
\label{wdbz004}
\end{figure}

\begin{figure}
\epsfig{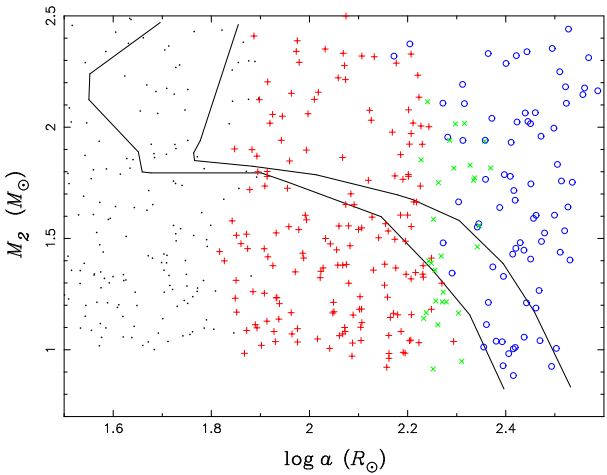}
\caption{
The distribution of WD binaries in the $a$--$M_2$
plane for systems where the first mass-transfer phase is stable but
non-conservative (with $\alpha_{\rm RLOF}=0.5$; similar to
Figure~\ref{a-rlof} but with $Z=0.004$). The symbols indicate the mass of the
white dwarf (as in Fig.~\ref{wdbz004}).
}
\label{a-rlofz4}
\end{figure}

We use the observations of Maxted et al.\ (2001) and Morales-Rueda et
al.\ (2002a,b) as our main data set to constrain the BPS
model. Observationally, two parameters of sdB binaries can be measured
accurately, the orbital period $P$ and the mass function $f$ which
just depends on the radial velocity amplitude. The latter can be
related to a minimum mass of the companion $M^{\rm min}_{\rm comp}$ by
choosing an inclination of $\sin i =1$ and adopting a typical mass for
the sdB star ($0.5\,M_\odot$ in the following). Since the minimum
companion mass is more closely related to the physical parameters of
the system, we follow common convention and plot the distribution of
both the observational data as well as our theoretical distributions
in a $M^{\rm min}_{\rm comp}$--$P$ diagram. In Figure~1 the filled
symbols show the distribution of the systems in the sample of
Maxted et al.\ (2001) and Morales-Rueda et al.\ (2002a,b), where
we excluded systems with known MS companions.

Since the majority of the sdB stars with known companions have WD companions
and since the orbital periods are less than 10\,d, this immediately
suggests that the majority of sdB binaries in this sample formed
through the second CE ejection channel (the second stable RLOF would
only produce sdB binaries with long orbital periods, $\sim
1000\,$d). Note that the first CE ejection channel also contributes
to the observational data set (in this case, the companion
is a main-sequence star instead of a white dwarf).

To illustrate how we can use this diagram as a diagnostic, we have
constructed a theoretical distribution of systems assuming that all
sdB binaries in our sample originate from the second CE ejection
channel.  For this purpose we use a simplified BPS model where we
adopt simple distributions for the systems before the second CE
phase. Specifically we assume here that the WD masses are uniformly
distributed between 0.25 and $0.45\,M_\odot$ and between 0.55 and
$0.65\,M_\odot$, that the mass of the sdB star progenitor on the main
sequence follows the IMF of Miller \& Scalo \shortcite{mil79}, and
that the logarithm of the separation, $\log (a/R_\odot)$, is uniformly
distributed between 1 and 4. We then determine the post-CE
parameters of the systems using our BPS code for chosen CE ejection
parameters $\alpha_{\rm CE}$ and $\alpha_{\rm th}$.  We further assume
that the normal directions of the orbital planes of the sdB stars are
randomly distributed and take the mass of the sdB star to be
$0.5\,M_\odot$, no matter what the actual mass in the simulation is, in 
order to get $M_{\rm WD}^{\rm min}$ which can then be compared directly
to the observational data set. 

In Figure~\ref{p-mwd} we plot the distribution of sdB stars resulting
from this simulation, where the symbols indicate the WD masses in the
simulation (dots: $0.25\leq M_{\rm WD}\leq 0.35\,M_\odot$, pluses:
$0.35< M_{\rm WD}\leq 0.45\,M_\odot$, circles: $0.55\leq M_{\rm
WD}\leq 0.65\,M_\odot$).  It is apparent that this simulation maps out
the observed range of the distribution reasonably well except for KPD
1930+2752, which has a WD mass of $0.97\,M_\odot$ (i.e. is more
massive than the white dwarfs in this simulation).  In this particular
simulation, the common-envelope ejection efficiency $\alpha_{\rm CE}$
and the thermal contribution to the CE ejection $\alpha_{\rm th}$ were
taken to be 0.75 (as in our best-fit model obtained in \S~7.4). We
have also tested lower and higher values for $\alpha_{\rm CE}$ and
$\alpha_{\rm th}$.  As one may imagine, higher values extend the
distribution further at long orbital periods and lower values limit
the distribution towards shorter orbital periods.

There is a small gap in the left part of the distribution. Subdwarf B
stars to the right of the gap are produced from systems where the ZAMS
mass of the progenitor is below the helium flash mass (i.e. the first
sub-channel), while sdB stars to the left had more massive ZAMS
progenitors (see \S~3.3).  The helium flash mass for Pop I is
$M_0=1.99\,M_\odot$. To allow a better interpolation in this mass
range, our model grid includes models with masses very close to the
helium flash mass with $M_{\rm ZAMS}=1.90\,M_\odot$ and $M_{\rm
ZAMS}=2.05\,M_\odot$, respectively.  If the masses of the two sets
were infinitesimally close to the helium flash mass, the gap would
disappear. However, this region would still be less densely populated
than neighbouring regions.

In order to understand the evolutionary history of these systems
better it is more instructive to look at the distribution of the
orbital separation $a$ and the mass of the progenitor of the sdB star,
$M_2$, for systems that become sdB binaries {\em before} the CE
phase. Figure~\ref{wdbinary} shows this distribution for the systems
shown in Figure~1, where the solid curves mark the boundary of the
parameters space that leads to the formation of short-period sdB
binaries. Identifying their evolutionary past then becomes a question
of what previous evolutionary paths will fill this particular region
of parameter space. This depends particularly on whether the first
mass-transfer phase, which leads to the formation of the white dwarf,
is dynamically stable or unstable.

In Figures~\ref{not-ce} and \ref{non-cons} we plot the distributions of
WD binaries after the first RLOF phase where the first mass-transfer
phase was unstable and stable, respectively (these were obtained from
BPS simulations with our standard set of assumptions; see \S~4.2).

From Figure~\ref{not-ce} it becomes immediately clear that systems
where the first mass-transfer phase is dynamically unstable and leads
to a CE phase are not likely to be responsible for the production of
WD binaries with the required parameters. For the case $\alpha_{\rm
CE}=0.75$ and $\alpha_{\rm th}=0.75$ (see case shown), only a few WD
binaries populate the marked region in the $a$--$M_2$ parameter
plane. We have tested that for values $\alpha_{\rm CE}=0.70$ and
$\alpha_{\rm th}=0.70$, no systems would satisfy this constraint.
Even in the most extreme case with $\alpha_{\rm CE}=1.0$ and
$\alpha_{\rm th}=1.0$, the maximum values physically allowed, the
right part of the marked region (with $\log a > 2.2$) which is in fact
the most important part, cannot be populated.  We can therefore safely
conclude that the first RLOF phase for the progenitors of short-period
sdB stars is likely to have been stable.

In Figure~\ref{non-cons} we plot the WD binaries that result from a
first stable RLOF phase assuming that mass transfer is conservative
(i.e. $\alpha_{\rm RLOF}=1$) and where we use $q_{\rm crit}=1.2$ in
the stability criterion (consistent with the results from Paper
I). The region of interest is now populated by two distinct groups of
systems separated by a gap. The WD binaries in the upper-left corner
are systems that experienced stable RLOF in the Hertzsprung gap while
systems below the gap fill their Roche lobe first on the
FGB\footnote{The gap is partly due to the fact that the radius of a
star shrinks near the end of the Hertzsprung gap; hence the core mass
for stars filling their Roche lobes on the FGB is somewhat larger
than at the end of the gap. Moreover, the size of the gap is also
determined by the definition of the core mass. As part of the envelope
mass is lost from the system, a large envelope mass (or a small core
mass) means that more angular momentum is lost during the stable RLOF
phase leading to a smaller separation.}.  This evolutionary path tends
to produce low-mass white dwarfs ($\la 0.35\,M_\odot$, indicated as
dots), so this cannot explain the many more massive WDs in
Figure~\ref{p-mwd}.  This implies that the first RLOF phase cannot be
conservative, at least not as a rule.

In Figure~\ref{a-rlof} we show a similar distribution, but now
assuming that the first RLOF is non-conservative with $\alpha_{\rm
RLOF}=0.5$ and that the mass lost takes away the same specific angular
momentum as pertains to the system ($\alpha=1$ in the PJH
formalism). The distribution fills the parameter space of interest
reasonably well, and the WD masses are also widely distributed as
required.  We generally find that using lower values of $\alpha_{\rm
RLOF}$, reduces the mass $M_2$ and shortens the orbital period.  If
$\alpha_{\rm RLOF}$ is too small (i.e. mass transfer is very
non-conservative), the most important part of the parameter space, the
lower-right part with $\log a > 2$, cannot be filled.  Increasing the
value of $\alpha$ increases the angular-momentum loss per unit mass
lost from the system.  Hence higher values of $\alpha$ produce shorter
orbital periods.  Again the lower-right part cannot be filled for
values of $\alpha =1.5$ and larger. For $\alpha=0.5$, all the parts of
the space are filled, but only with relatively low-mass WDs
($\la 0.45\,M_\odot$).  We also tested some cases where the mass lost takes
away the same specific angular momentum as pertains to the mass donor
or the mass gainer (for $\alpha_{\rm RLOF}=0.75$, 0.50 and 0.25,
respectively). In these cases, all parts of the parameter space are
filled, but only with low-mass white dwarfs ($<0.45\,M_\odot$). 

The main conclusion of these comparisons is that, in order to obtain a
wide coverage of the parameter space that can lead to the formation of
short-period sdB binaries, {\em the first phase of mass transfer has
to be non-conservative}, where our best-choice parameters are
$\alpha_{\rm RLOF}=0.5$ and $\alpha \simeq 1.0$.

All of these results are, however, dependent on the metallicity of the
population.  To examine this, we carried out a similar set of tests for a
typical thick-disc metallicity of $Z=0.004$. The results of these
simulations are shown in Figures~\ref{pmwdz004} to \ref{a-rlofz4}.  The
results are broadly similar, except that there is a systematic shift
in the distribution towards shorter separations and lower masses $M_2$
(most clearly seen when comparing Figs.~5 and 8).

Finally we note that, if we had used the criterion for stable RLOF
based on a polytropic model \cite{hje87,web88,sob97,han01}, revised
to take into account non-conservative mass transfer, we
would have obtained a very small number of WD binaries, but none
of them would actually populate the required parameter space
for $\alpha_{\rm RLOF}=0.5$  and $\alpha =1.0$.

\section{Monte Carlo simulations}

\begin{table*}
 \caption{Birthrates of sdB stars from different channels 
(in $10^{-3}\,{\rm yr}^{-1}$)}
 \begin{tabular}{lllllllllllll}
 \hline\hline
set & $Z$ & $n(1/q)$ & $q_{\rm crit}$ & $\alpha_{\rm CE}$ & $\alpha_{\rm th}$ &
 1st CE & 1st RLOF & 2nd CE & 2nd RLOF & sdB Binary & Merger & Total\\
 \hline
& &&&&&&&&&&\\
1&0.02 & a & 1.5 & 0.5 & 0.5 &
    5.51&   29.89&    2.80&    0.00&   38.20&   17.22&   55.42 \\
2& 0.02 & a & 1.5 & 0.75 & 0.75 &
    6.80&   29.89&    5.44&    0.00&   42.13&   16.62&   58.75 \\
3& 0.02 & a & 1.5 & 1.0 & 1.0 &
    8.41&   29.89&    8.38&    0.00&   46.68&   16.24&   62.93 \\
& &&&&&&&&&&\\
4& 0.02 & b & 1.5 & 0.5 & 0.5 &
    7.22&    3.46&    0.32&    0.00&   11.00&    3.30&   14.31 \\
5& 0.02 & b & 1.5 & 0.75 & 0.75 &
    9.16&    3.46&    0.55&    0.00&   13.17&    3.22&   16.39 \\
6& 0.02 & b & 1.5 & 1.0 & 1.0 &
   11.23&    3.46&    0.79&    0.00&   15.48&    3.06&   18.54 \\
& &&&&&&&&&&\\
7& 0.02 & a & 1.2 & 0.5 & 0.5 &
    7.02&   22.25&    1.58&    0.00&   30.84&    8.51&   39.36 \\
8& 0.02 & a & 1.2 & 0.75 & 0.75 &
    8.62&   22.25&    2.98&    0.00&   33.85&    8.25&   42.09 \\
9& 0.02 & a & 1.2 & 1.0 & 1.0 &
   10.71&   22.25&    5.43&    0.00&   38.38&    7.99&   46.38 \\
& &&&&&&&&&&\\
10& 0.004 & a & 1.2 & 0.5 & 0.5 &
    8.21&   26.22&    2.00&    0.00&   36.43&   10.28&   46.71 \\
11& 0.004 & a & 1.2 & 0.75 & 0.75 &
   10.56&   26.22&    3.82&    0.00&   40.60&    9.95&   50.55 \\
12& 0.004 & a & 1.2 & 1.0 & 1.0 &
   13.19&   26.22&    5.79&    0.00&   45.20&    9.38&   54.58 \\
 \hline
 \end{tabular}

 \medskip
 \label{birthrate}
\end{table*}

\begin{table*}
 \caption{Percentages of sdB stars from different channels and the total
numbers (in $10^6$) in the Galaxy at the current epoch}
 \begin{tabular}{lllllllllllll}
 \hline\hline
 set & $Z$ & $n(1/q)$ & $q_{\rm crit}$ & $\alpha_{\rm CE}$ & $\alpha_{\rm th}$ &
 1st CE & 1st RLOF & 2nd CE & 2nd RLOF & sdB Binary & Merger & Total Number\\
 &&&&&&&&&&&&($10^6$)\\
 \hline
 &&&&&&&&&&&\\
 1&0.02 & a & 1.5 & 0.5 & 0.5 &
  14.63&  61.75&   4.88&   0.00&  81.25&  18.75&   7.04 \\
  &&&&&&
  28.09&   0.00&  14.85&   0.00&  42.94&  57.06&   2.31 \\
{\vspace{3pt}}
  &&&&&& 
  53.72&   0.00&  22.37&   0.00&  76.09&  23.91&   1.08 \\
 2&0.02 & a & 1.5 & 0.75 & 0.75 &
  17.92&  55.05&   5.08&   0.00&  78.05&  21.95&   7.92 \\
  &&&&&&
  27.93&   0.00&  13.55&   0.00&  41.49&  58.51&   2.97 \\
{\vspace{3pt}}
  &&&&&&
  39.35&   0.00&  15.65&   0.00&  55.00&  45.00&   1.63 \\
 3&0.02 & a & 1.5 & 1.0 & 1.0 &
  19.74&  45.56&  10.63&   0.00&  75.94&  24.06&   9.52 \\
  &&&&&&
  24.26&   0.00&  23.21&   0.00&  47.47&  52.53&   4.36 \\
  &&&&&&
  27.51&   0.00&  17.24&   0.00&  44.75&  55.25&   2.45 \\
 &&&&&&&&&&&\\
 4&0.02 & b & 1.5 & 0.5 & 0.5 &
  61.32&  25.46&   2.30&   0.00&  89.08&  10.92&   2.41 \\
  &&&&&&
  80.68&   0.00&   3.36&   0.00&  84.04&  15.96&   1.65 \\
{\vspace{3pt}}
  &&&&&&
  92.96&   0.00&   2.95&   0.00&  95.91&   4.09&   1.24 \\
 5&0.02 & b & 1.5 & 0.75 & 0.75 &
  67.23&  19.42&   1.93&   0.00&  88.59&  11.41&   3.15 \\
  &&&&&&
  81.72&   0.00&   2.65&   0.00&  84.37&  15.63&   2.30 \\
{\vspace{3pt}}
  &&&&&&
  87.96&   0.00&   2.56&   0.00&  90.52&   9.48&   1.58 \\
 6&0.02 & b & 1.5 & 1.0 & 1.0 &
  70.34&  15.37&   3.32&   0.00&  89.04&  10.96&   4.07 \\
  &&&&&&
  81.45&   0.00&   4.32&   0.00&  85.76&  14.24&   3.14 \\
  &&&&&&
  81.77&   0.00&   2.98&   0.00&  84.75&  15.25&   1.76 \\
 &&&&&&&&&&&\\
 7&0.02 & a & 1.2 & 0.5 & 0.5 &
  30.82&  46.19&   5.49&   0.00&  82.51&  17.49&   4.12 \\
  &&&&&&
  40.26&   0.00&  14.28&   0.00&  54.54&  45.46&   1.58 \\
{\vspace{3pt}}
  &&&&&&
  63.83&   0.00&  20.24&   0.00&  84.07&  15.93&   0.87 \\
 8&0.02 & a & 1.2 & 0.75 & 0.75 &
  36.37&  39.44&   5.17&   0.00&  80.98&  19.02&   4.80 \\
  &&&&&&
  41.18&   0.00&  12.57&   0.00&  53.76&  46.24&   1.97 \\
{\vspace{3pt}}
  &&&&&&
  52.50&   0.00&  14.40&   0.00&  66.90&  33.10&   1.18 \\
 9&0.02 & a & 1.2 & 1.0 & 1.0 &
  37.73&  31.31&  11.68&   0.00&  80.72&  19.28&   6.04 \\
  &&&&&&
  35.93&   0.00&  24.18&   0.00&  60.11&  39.89&   2.92 \\
  &&&&&&
  39.97&   0.00&  16.92&   0.00&  56.89&  43.11&   1.63 \\
 &&&&&&&&&&&\\
 10&0.004 & a & 1.2 & 0.5 & 0.5 &
  31.51&  44.40&   6.16&   0.00&  82.07&  17.93&   4.33 \\
  &&&&&&
  37.10&   0.00&  16.08&   0.00&  53.17&  46.83&   1.66 \\
{\vspace{3pt}}
  &&&&&&
  59.05&   0.00&  23.59&   0.00&  82.64&  17.36&   0.95 \\
 11&0.004 & a & 1.2 & 0.75 & 0.75 &
  36.33&  36.57&   6.22&   0.00&  79.12&  20.88&   5.16 \\
  &&&&&&
  34.95&   0.00&  14.93&   0.00&  49.88&  50.12&   2.15 \\
{\vspace{3pt}}
  &&&&&&
  46.55&   0.00&  16.93&   0.00&  63.47&  36.53&   1.34 \\
 12&0.004 & a & 1.2 & 1.0 & 1.0 &
  38.35&  28.56&  12.70&   0.00&  79.62&  20.38&   6.66 \\
  &&&&&&
  31.44&   0.00&  26.32&   0.00&  57.76&  42.24&   3.22 \\
  &&&&&&
  36.96&   0.00&  17.03&   0.00&  53.99&  46.01&   1.82 \\
 \hline
 \end{tabular}

 \medskip
 \label{percentage}
\end{table*}

In order to investigate the formation of sdB stars from the various
channels more systematically, we performed 12 sets of Monte Carlo
simulations altogether for a Pop I and a thick-disc population
($Z=0.004$) by varying the model parameters over a reasonable range.
Specifically, we varied the parameter $\alpha_{\rm CE}$ for the CE
ejection efficiency and the parameter $\alpha_{\rm th}$ for the
thermal contribution to CE ejection from 0.5 to 1.0, the value of
$q_{\rm crit}$ in the criterion for a first phase of stable RLOF on
the FGB or AGB from 1.2 to 1.5. Two initial mass-ratio distributions
were adopted: a constant mass-ratio distribution and one where the
masses are uncorrelated and drawn independently from a Miller-Scalo
IMF.  Guided by the results from the previous section, we assume in
all of these simulations that the first stable RLOF phase is
non-conservative (with $\alpha_{\rm RLOF}=0.5$) and that the mass lost
takes away the same specific angular momentum as pertains to the
system.  We assume that one binary with its primary more massive than
$0.8\,M_\odot$ is formed annually in the Galaxy for both the Pop I and
the thick-disc population.  Note that this star-formation rate is
almost certainly too high for the thick-disc population and that
therefore these results should be scaled down accordingly.

Table~\ref{birthrate} lists the birthrates of sdB stars produced from
the various formation channels.  In the table, the 2nd column denotes
the metallicity ($Z=0.02$ for Pop I and $Z=0.004$ for the thick-disc
population); the 3rd column indicates the initial mass-ratio
distribution, where `a' represents a constant mass-ratio distribution and `b'
one of uncorrelated component masses; the 4th column gives $q_{\rm
crit}$, the critical mass ratio for the first stable RLOF on the FGB
or AGB; the 5th and the 6th columns give the values of $\alpha_{\rm
CE}$ and $\alpha_{\rm th}$ adopted, respectively.  Galactic birthrates
for sdB stars (in $10^{-3}\,{\rm yr}^{-1}$) from the first CE ejection
channel, the first stable RLOF channel, the second CE ejection channel
and the second stable RLOF channel are listed in columns 7 to 10.  The
3rd column from the right gives the birthrates of sdB binaries, and
the 2nd column from the right gives the birthrates of single sdB stars
resulting from the helium WD merger channel. The last column
gives the total birthrates of sdB stars from all channels.

Table~\ref{percentage} lists the percentages of sdB stars from various
channels and the total numbers in the Galaxy at the current epoch.
Columns 1 - 6 list the main model parameters as in
Table~\ref{birthrate}. Percentages of sdB stars from the first CE
ejection channel, the first stable RLOF channel, the second CE
ejection channel and the second stable RLOF channel are given in
columns 7 to 10.  The 3rd column from the right gives the percentages
of sdB binaries, and the 2nd column from the right the percentages of
single sdB stars resulting from the helium WD merger channel. The last
column gives the total numbers (in $10^6$) of sdB stars from all the
channels in the Galaxy.

For each item in the table we list three numbers. The first row
gives the number for sdB stars without taking any observational
selection effects into account and therefore represents
the intrinsic distribution, the second row takes into account
the GK selection effect, i.e. excludes any sdB binaries where the
secondary is of spectral type K or earlier, and the third row takes into
account the GK and the strip selection effects as to best
represent the sample of Maxted et al.\ (2001).

Various model parameters from these simulations are plotted
in Figures~\ref{period} to ~\ref{theta}, which will be discussed in
detail in the next section.

\section{Discussion}

As summarized in \S~3, we altogether consider five channels for the
formation of sdB stars: the first CE ejection channel, the first
stable RLOF channel, the second CE ejection channel, the second stable
RLOF channel and the double He WD merger channel.  The birthrates of sdB
stars formed from each channel and the relative percentages at the
current epoch are listed in Tables~\ref{birthrate} and
~\ref{percentage}, respectively.

As these tables show, the  relative  importance of the five channel
varies significantly for the different sets of parameters. 
The first CE ejection, the first stable RLOF and the merger channels
are the most important ones intrinsically. However, once selection effects
are taken into account, the second CE ejection channel becomes of comparable
importance. As mentioned before, for our set of assumptions we do not obtain 
sdB stars from the second stable RLOF channel. Note also that the GK
selection effect tends to remove all of the sdB binaries from the first 
stable RLOF channel.


\subsection{Sensitivity to the model parameters}

Our BPS model requires a number of model parameters and input
distributions. The parameters/distributions which we varied in the study
are: $q_{\rm crit}$, the critical mass ratio above which mass transfer
is dynamically unstable on the FGB/AGB, the mass-transfer efficiency
$\alpha_{\rm RLOF}$, which defines the fraction of mass lost from the
primary that is accreted by the gainer for systems experiencing stable
RLOF after the main-sequence phase, the specific angular momentum
$\alpha$ lost from the system/unit mass during stable RLOF, the CE
ejection efficiency $\alpha_{\rm CE}$ and the thermal contribution
factor $\alpha_{\rm th}$ in the CE ejection criterion, the initial
mass-ratio distribution $n(1/q)$ and the metallicity of the
population.

As was shown in \S~5, $q_{\rm crit}$, $\alpha_{\rm RLOF}$ and $\alpha$
are strongly constrained by the $\log P$ - $M_{\rm WD}^{\rm min}$
diagram of the observations by Maxted et al.\ (2001) and Morales-Rueda
et al.\ (2002a,b). In order to match the observed
distribution, the value for $q_{\rm crit}$ cannot be taken from a
simple polytropic model
\cite{hje87,web88,sob97}, even in a revised version taking
non-conservative RLOF into account
\cite{han01}, as such a $q_{\rm crit}$ would make a first phase of stable RLOF
very unlikely and would not produce WD binaries with the parameters
required to explain the sample of Maxted et al.\ (2001).  Completely
conservative RLOF ($\alpha_{\rm RLOF}=1$) or the assumption that the
mass lost from the system takes away the same specific angular
momentum as pertains to the primary/secondary also cannot explain the
observations. This analysis favours values $\alpha_{\rm RLOF}\simeq
0.5$ and $\alpha \simeq 1$ (in units of ${2\pi a^2/P}$); we adopted
these value for all of our simulations.

We investigated two values for $q_{\rm crit}$, 1.2 and 1.5.  The higher
value implies that the mass donor can be more massive and that
the first stable RLOF phase then results in WD binaries with more
massive WD companions (see Figures~\ref{pm1} and ~\ref{pm2}). 
Obviously, a higher
$q_{\rm crit}$ leads to fewer sdB stars from the first CE ejection
channel, more from the first stable RLOF channel and more from the second
CE ejection channel.  As a consequence, the birthrate of sdB binaries is
increased significantly (see Table~\ref{birthrate}); however, the
fraction of sdB binaries is not influenced significantly, as the
merger rate also increases.

An increase in $\alpha_{\rm CE}$ and $\alpha_{\rm th}$ makes it easier
to eject the CE and hence leads to a systematic increase in the
post-CE orbital periods of sdB binaries from the first CE ejection and
the second CE ejection; it also leads to higher birthrates from
these two channels, but decreases the rate from the merger channel
(since fewer systems will merge in the age of the Galaxy). However,
the binary fraction of sdB stars at the current epoch decreases. The
reason is that the envelope of a star near the tip of the FGB for ZAMS
masses less than the helium flash mass $M_0$ is loosely bound and can
be easily ejected for a wide range of these parameters.
Therefore the main effect of an increase in $\alpha_{\rm CE}$ and
$\alpha_{\rm th}$ is to increase the numbers of CE ejections for stars
with ZAMS masses greater than $M_0$. As their envelopes are tightly
bound and the sdB binaries formed this way have very short orbital
periods, they merge soon after their formation. On the other hand, the
increase of $\alpha_{\rm CE}$ and $\alpha_{\rm th}$ makes helium WD
pairs with a low total mass more likely, and therefore the sdB stars
from the merger channel generally have a lower mass. Since the
timescale for helium burning for a low-mass sdB star is significantly
longer in this case, this leads to an increased contribution of sdB
stars formed through the merger channel at the current epoch.

As compared to the constant initial mass-ratio distribution,
the distribution for uncorrelated component masses means that
a star is more likely to have a low-mass companion. Therefore this
distribution leads to more sdB stars from the first CE ejection
channel and greatly decreases the numbers of sdB stars from the
first stable RLOF, the second CE ejection and the merger channel.
The overall result is that the binary fraction of sdB stars increases 
significantly by about 10 per cent. 

The evolutionary timescale for stars of a given mass is shorter for
stars with a thick-disc metallicity $Z=0.004$ than the corresponding
timescale for Pop I objects. This implies that stars of lower mass can
evolve to become sdB star within the age of the Galaxy. Since these
lower-mass stars are relatively more common, the numbers of sdB stars
of a thick-disc population from all the channels would be larger than
that of Pop I for the same star-formation rate.  The binary fraction
of sdB stars at the current epoch is somewhat higher than for Pop I.
The reason is that sdB binaries from the CE ejection channels are more
likely from low-mass FGB stars for a thick-disc population due to
their shorter evolutionary timescales, and the orbital periods of
those sdB binaries resulting from low-mass FGB stars are so long that
they will not merge in the lifetime of the Galaxy. A higher WD mass is
also more likely for sdB binaries for the thick-disc population from
the second CE ejection channel (see Fig.~\ref{pm2}), as the core
mass of a FGB/AGB star is more likely to be massive and therefore the
WD binary resulting from the first RLOF is more likely to contain a
massive WD.
 
\subsection{The distribution of orbital periods}

\begin{figure*}
\epsfig{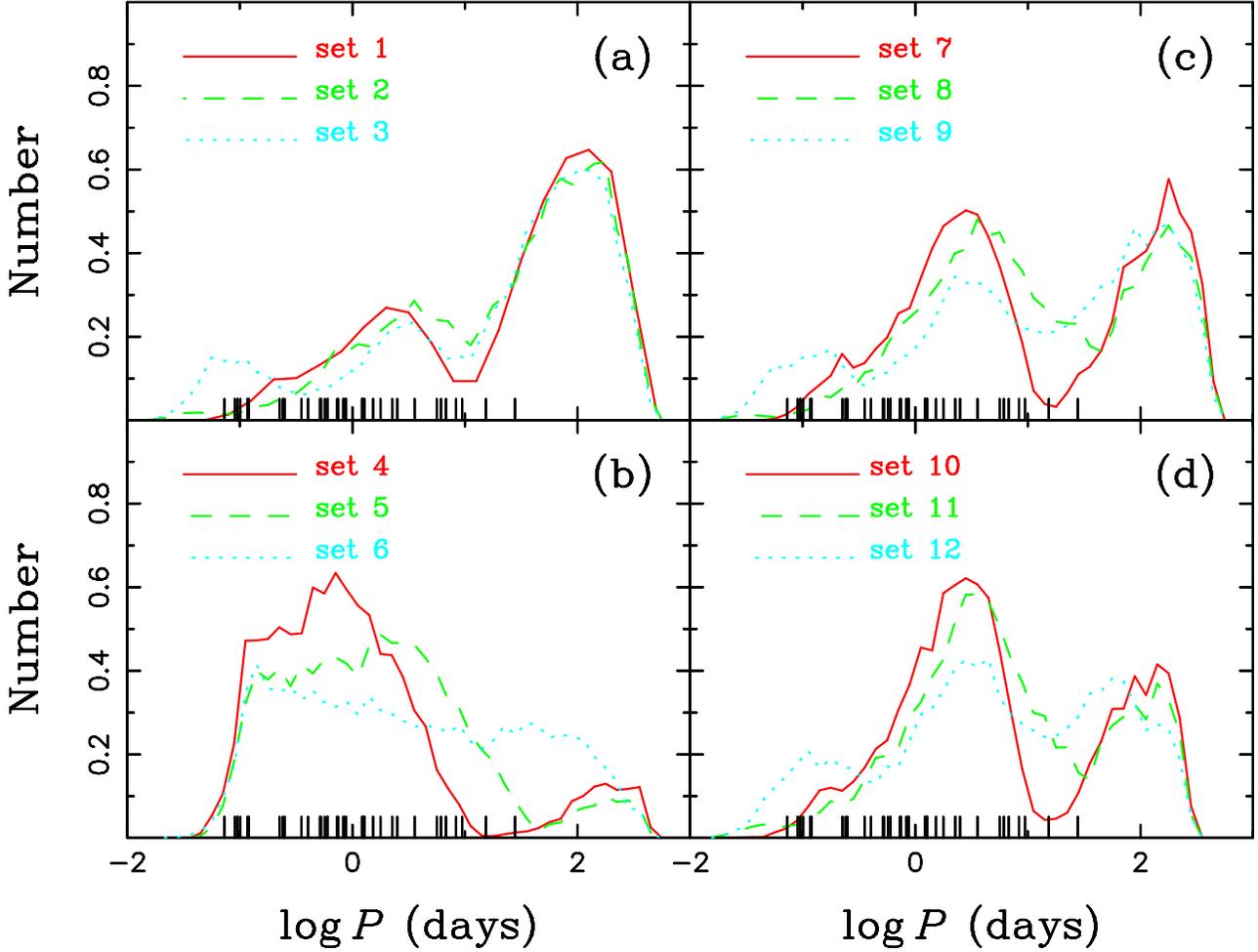}
\caption{
The distribution of orbital periods of sdB stars from all the
simulation sets.  The short ticks along the X-axis indicate the
positions of sdB stars in the observational sample (Maxted et al.\
2001; Morales-Rueda et al.\ 2002a,b).  Panels (a) and (b) illustrate
how the results depend on the initial mass ratio distribution with
panel (a) using a flat distribution and panel (b) a distribution with
uncorrelated component masses. Panels (a) (with $q_{\rm crit}=1.5$)
and (c) (with $q_{\rm crit}=1.2$) show the effect of changing the
critical mass ratio for stable RLOF on the FGB/AGB. Panels (c) and (d)
demonstrate the metallicity dependence (with $Z=0.02$ and 0.004,
respectively).  }
\label{period}
\end{figure*}

\begin{figure}
\epsfig{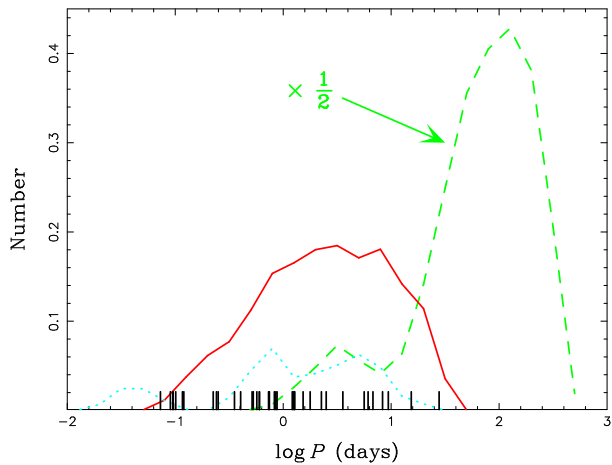}
\caption{
The distribution of orbital periods of sdB stars from different
channels in simulation set 2 
(with $Z=0.02$, a flat mass ratio distribution,
$q_{\rm crit}=1.5$, $\alpha_{\rm CE}=\alpha_{\rm th}=0.75$)
(solid: the first CE ejection channel,
dashed: the first stable RLOF channel, dotted: the second CE ejection
channel).  No sdB stars are produced from the second stable RLOF
channel.  The curve for the first stable RLOF channel has been
rescaled by a factor of 1/2 for clarity. Short ticks along the X-axis
indicate the positions of sdB stars in the observational sample
(Maxted et al.\ 2001; Morales-Rueda et al.\ 2002a,b.}
\label{p-chan}
\end{figure}

Figure~\ref{period} shows the distribution of orbital periods of sdB
binaries at the current epoch for simulation sets 1 to 12.  The
orbital period ranges from 0.5\,hr to 500\,d (see also
Figure~\ref{p-chan}). While the lower limit is essentially fixed by
the condition that neither component fills its Roche lobe and
therefore mainly depends on the radii of both components, the upper
limit is mainly determined by how much angular momentum is lost during
the first stable RLOF phase.  Some of the distributions, such as those
in simulation sets 3, 9 and 12 with high CE ejection efficiency, have
three peaks. The leftmost peak comes from the second CE ejection
channel where the donor's ZAMS mass is greater than the helium flash
mass $M_0$. In this case, the envelope is tightly bound and the WD has
to penetrate very deep before the envelope can be ejected, which leads
to a very short orbital period of the sdB binary.  The central peak
contains systems from the first and the second CE ejection channels
where the donor's ZAMS mass is less than $M_0$. The envelope is more
loosely bound in this case, leading to a longer post-CE orbital
period. The rightmost peak is due to sdB binaries from the first
stable RLOF channel, which always produces systems with long orbital
periods.

Panel (b) shows that sdB binaries from the first CE ejection channel
dominate for the simulations where the initial component masses are
uncorrelated. Note that in this case very few sdB binaries are formed
in the first RLOF channel since for uncorrelated masses the first
mass-transfer phase is dynamically unstable in most cases.  

In Figure~\ref{p-chan} we present the distribution of orbital periods
of sdB binaries at the current epoch from the different channels of
simulation set 2 (with $Z=0.02$, a flat mass ratio distribution,
$q_{\rm crit}=1.5$, $\alpha_{\rm CE}=\alpha_{\rm th}=0.75$). 
sdB stars from the first CE ejection channel have
orbital periods from 1.5\,hr to 40\,d, where the lower limit is again
constrained by the radii of the MS companions and the upper limit
depends strongly on the CE ejection efficiency $\alpha_{\rm CE}$ and
the thermal contribution $\alpha_{\rm th}$ to the CE ejection. In the
extreme case ($\alpha_{\rm CE}=\alpha_{\rm th}=1$), the upper limit
can be as high as 400\,d. sdB stars from the first stable RLOF channel
have orbital periods from 15\,hr to 500\,d, and the orbital-period range is
sensitive to the assumption concerning the systemic angular momentum
loss during the first stable RLOF phase. The distribution has two
peaks.  The left peak is caused by sdB stars that experience stable
RLOF in the Hertzsprung gap where the donor's ZAMS mass is larger than
$M_0$, while the right peak is dominated by systems undergoing stable
RLOF on the FGB.  (Note that the minimum mass of the helium remnant
which will ignite helium in the core is $\sim 0.33\,M_\odot$ for stars
with a ZAMS mass greater than $M_0$; this value depends, however, on
the initial mass ratio; see Han, Tout \& Eggleton 2000 for
details.) The left peak is much smaller than the right peak since all
the donors in this group have to be quite massive and hence for the
adopted IMF have a lower probability. In addition, their companions
are likely to be massive as well in this simulation with a constant
mass-ratio distribution and have a relatively short evolution
time. For example, the lifetime of a Pop I star with a ZAMS mass of
$2.5\,M_\odot$ on the main sequence 
is $\sim 7.7\times 10^8\,{\rm yr}$ and the lifetime of a $3.2\,M_\odot$
star is $\sim 4.0\times 10^8\,{\rm yr}$.  Such lifetimes are comparable
to the core-helium burning lifetime of low-mass sdB stars: e.g. the
core-helium burning lifetime of a $0.35\,M_\odot$ sdB star is $\sim 6.2
\times 10^8 \,{\rm yr}$.  This has the consequence that the companion
star will fill its Roche lobe while the sdB star is still burning
helium in the core and the system may then no longer have the
appearance of a sdB binary (i.e. be a helium-burning star with a
{\em thin} hydrogen-rich envelope).
On the other hand, a donor experiencing stable RLOF on the FGB is
likely to be less massive and both components will have longer
evolutionary timescales and mass transfer will not occur while
the sdB star is still in the helium core-burning phase. 

The second CE ejection channel produces sdB binaries with orbital
periods from 0.5\,hr to 25\,d. The distribution has two parts,
separated by a well-defined gap.  The left part contains systems where
the donor's ZAMS mass is greater than $M_0$, and the right part
systems with ZAMS donor masses less than $M_0$. The gap is caused by
the sharp drop of the radius at the tip of the FGB from stars with
ZAMS masses somewhat smaller than $M_0$ relative to stars with ZAMS
masses somewhat greater than $M_0$ (see Fig.~8 and Table~1 of Paper
I).  This sharp drop leads to a great decrease in the radius range for
which a star can fill its Roche lobe, eject the common envelope and is
then able to ignite helium in the core. The right part has two peaks,
caused by the bimodal distribution of the masses of the WD
companions. Note that the sdB stars from the first CE ejection channel
fill in the gap because of the large range of masses for the MS
companions.  In contrast, sdB stars from the second CE ejection
channel all have WD companions whose masses are restricted to a rather
narrow range. The second stable RLOF channel does not produce any sdB
stars in this model since mass-transfer is dynamically unstable in all
cases. To obtain systems from this channel requires a tidally enhanced
stellar wind. As shown in Paper I, sdB binaries produced from the
second stable RLOF phase would have orbital periods in the range of
400 to 1500\,d.

\subsection{The distribution of masses}

\begin{figure*}
\epsfig{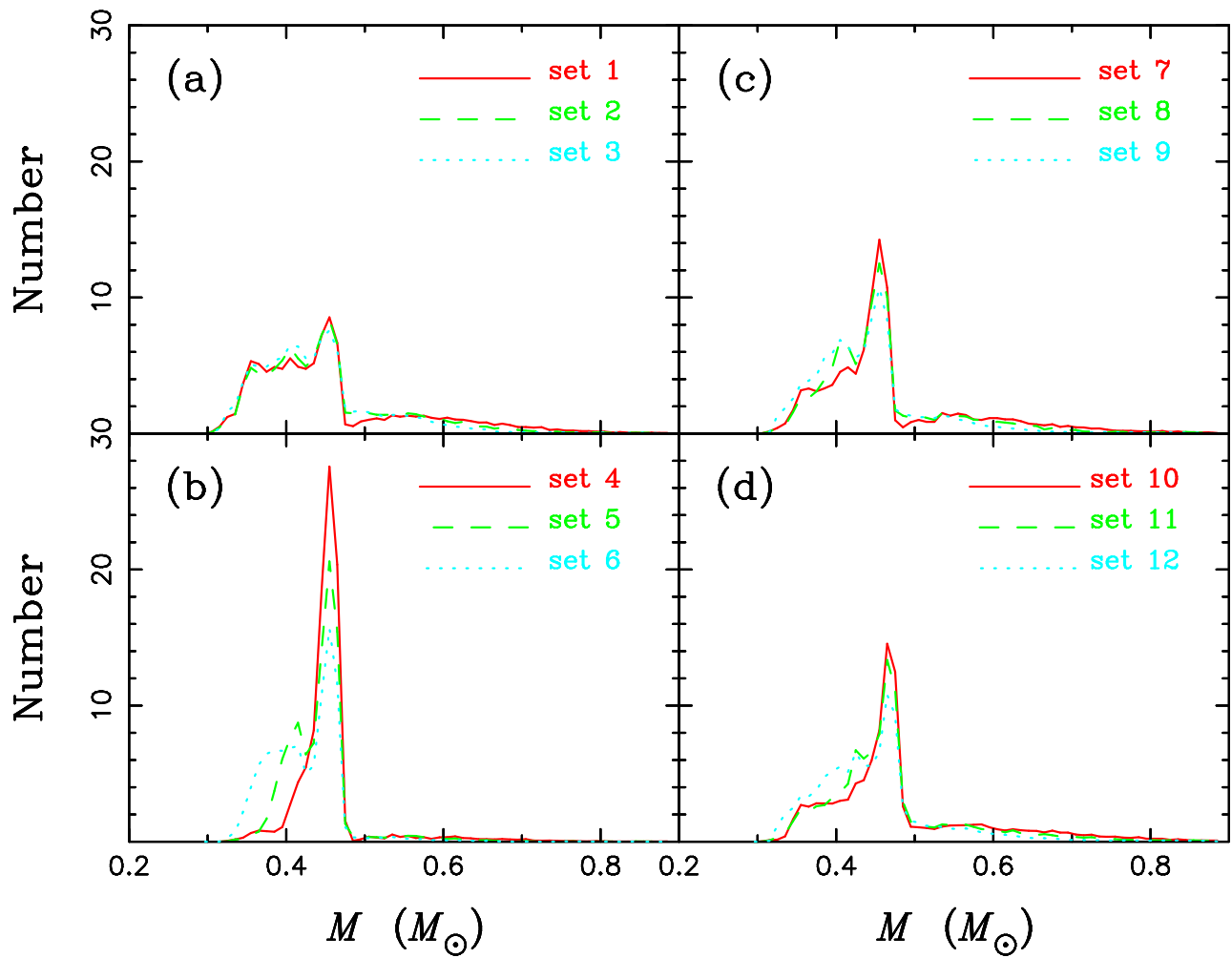}
\caption{
 Distribution of masses of sdB stars from all the simulation sets.
 Panels (a) and (b) show the effect of a change in the initial mass
 ratio distribution with panel (a) using a flat distribution and panel
 (b) a distribution of uncorrelated component masses. The effect of a
 change in the critical mass ratio for stable RLOF on the FGB/AGB is
 shown in panels (a) ($q_{\rm crit}=1.5$) and (c) ($q_{\rm
 crit}=1.2$). Panels (c) and (d) are for different metallicities
 ($Z=0.02$ and 0.004, respectively).  }
\label{mass}
\end{figure*}

\begin{figure}
\epsfig{file=m-chan.cps,angle=270,width=8cm}
\caption{
Distribution of masses of sdB stars from different channels in simulation
set 2 
(with $Z=0.02$, a flat mass ratio distribution,
$q_{\rm crit}=1.5$, $\alpha_{\rm CE}=\alpha_{\rm th}=0.75$)
(solid: the first CE ejection channel, dashed: the first 
stable RLOF channel, dot-dashed: the second CE ejection channel,
dotted: the merger channel).
No sdB stars are produced from the second stable RLOF in this simulation.
}
\label{m-chan}
\end{figure}

Figure~\ref{mass} displays the distributions of the masses of sdB
stars from all the simulation sets. The overall mass range is quite
wide, ranging from $\sim 0.3\,M_\odot$ to $\sim 0.8\,M_\odot$. The
distribution does not depend much on the CE ejection efficiency
($\alpha_{\rm CE}$) or the thermal contribution to CE ejection
($\alpha_{\rm th}$). The distribution is mainly determined by $q_{\rm
crit}$, the critical mass ratio for stable RLOF on the FGB or the AGB,
and the initial mass-ratio distribution.  As a matter of fact, the
distribution is mainly controlled by the contribution of systems from
the first stable RLOF channel. When this contribution is large, as in
simulation sets 1, 2 and 3 ($q_{\rm crit}=1.5$ and with a flat initial
mass ratio distribution), the distribution is wide and flat (0.35 -
$0.47\,M_\odot$ in panel a). If the contribution is insignificant, as
in simulation sets 4, 5 and 6 ($q_{\rm crit}=1.5$ and with an initial
mass ratio distribution of uncorrelated component masses), 
the distribution is narrow and sharply
peaked (the peak at $0.46\,M_\odot$ in panel b).

Figure~\ref{m-chan} gives the distributions from different channels in
simulation set 2
(with $Z=0.02$, a flat mass ratio distribution,
$q_{\rm crit}=1.5$, $\alpha_{\rm CE}=\alpha_{\rm th}=0.75$). 
The distribution for the 1st CE ejection channel has
a sharp major peak at $0.46\,M_\odot$ and a secondary peak at
$0.4\,M_\odot$. The secondary peak is due to the fact that the stellar
radius range for CE ejection near the tip of the FGB to produce sdB
stars is wider for $M_{\rm ZAMS}=1.90$ than for $M_{\rm
ZAMS}\le1.60\,M_\odot$ (see Table~1 in Paper I); as a consequence, CE
ejection for stars around $M_{\rm ZAMS}=1.90$ results in relatively
low-mass sdB stars which have relatively long helium-burning
lifetimes.

The first stable RLOF channel produces a plateau (or broad peak) in the
distribution at low masses, and the distribution drops sharply near
$0.47\,M_\odot$, as most systems experiencing stable RLOF in the
Hertzsprung gap result in low-mass sdB stars, while the maximum mass
is limited by the core mass at the tip of the FGB at which the helium
flash or helium ignition occurs.

The distribution for the 2nd CE ejection channel has three peaks, the
small left one at $0.33\,M_\odot$ represents systems with a ZAMS donor
mass greater than the helium flash mass. The central one and the right
one are analogous to the two peaks in the first CE ejection channel.
Finally, the merger channel produces a relatively wide and flat
distribution from 0.42 to $0.72\,M_\odot$.

\subsection{The best-choice model}

\begin{figure*}
\epsfig{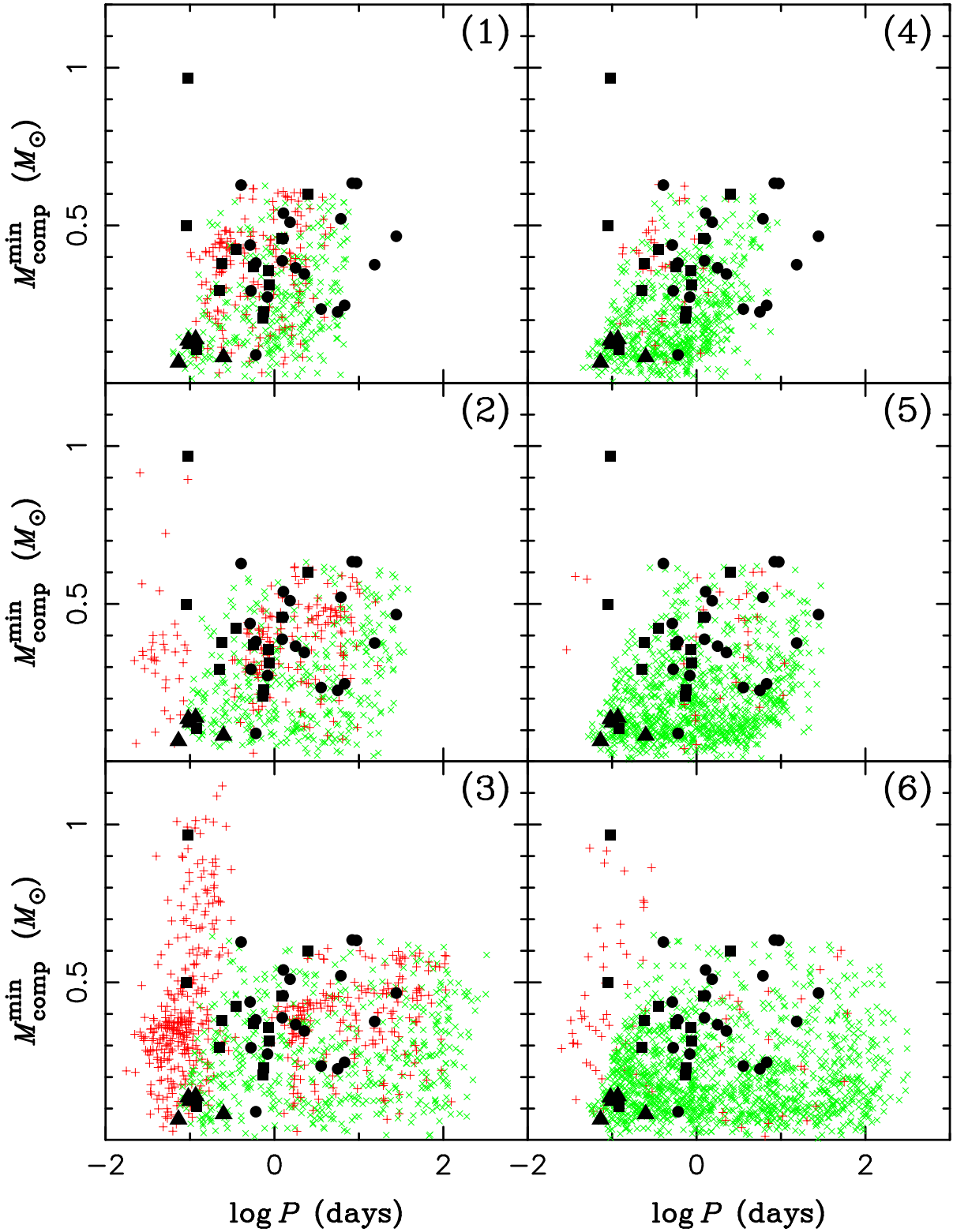}
\caption{
The distribution of sdB stars at the current epoch from simulations 1
to 6 (as indicated in the upper-right corner of each panel) in a $\log
P$ - $M_{\rm comp}^{\rm min}$ diagram, where $P$ is the orbital period
and $M_{\rm comp}^{\rm min}$ the minimum companion mass as defined in
the observations of Maxted et al.\ (2001).  Filled squares represent
observed sdB stars with WD companions, filled triangles observed sdB
stars with dM companions and filled circles observed sdB stars with
companions of unknown type (see Morales-Rueda et al.\ 2002b). Pluses
indicate sdB stars formed from the 2nd CE ejection channel with WD
companions, crosses represent sdB stars from the 1st RLOF phase
(either the first CE ejection or the first stable RLOF channel) with
MS companions.  In all panels, the GK selection is included
(i.e. systems with MS companions hotter than $4000$K were excluded).
}
\label{pm1}
\end{figure*}

\begin{figure*}
\epsfig{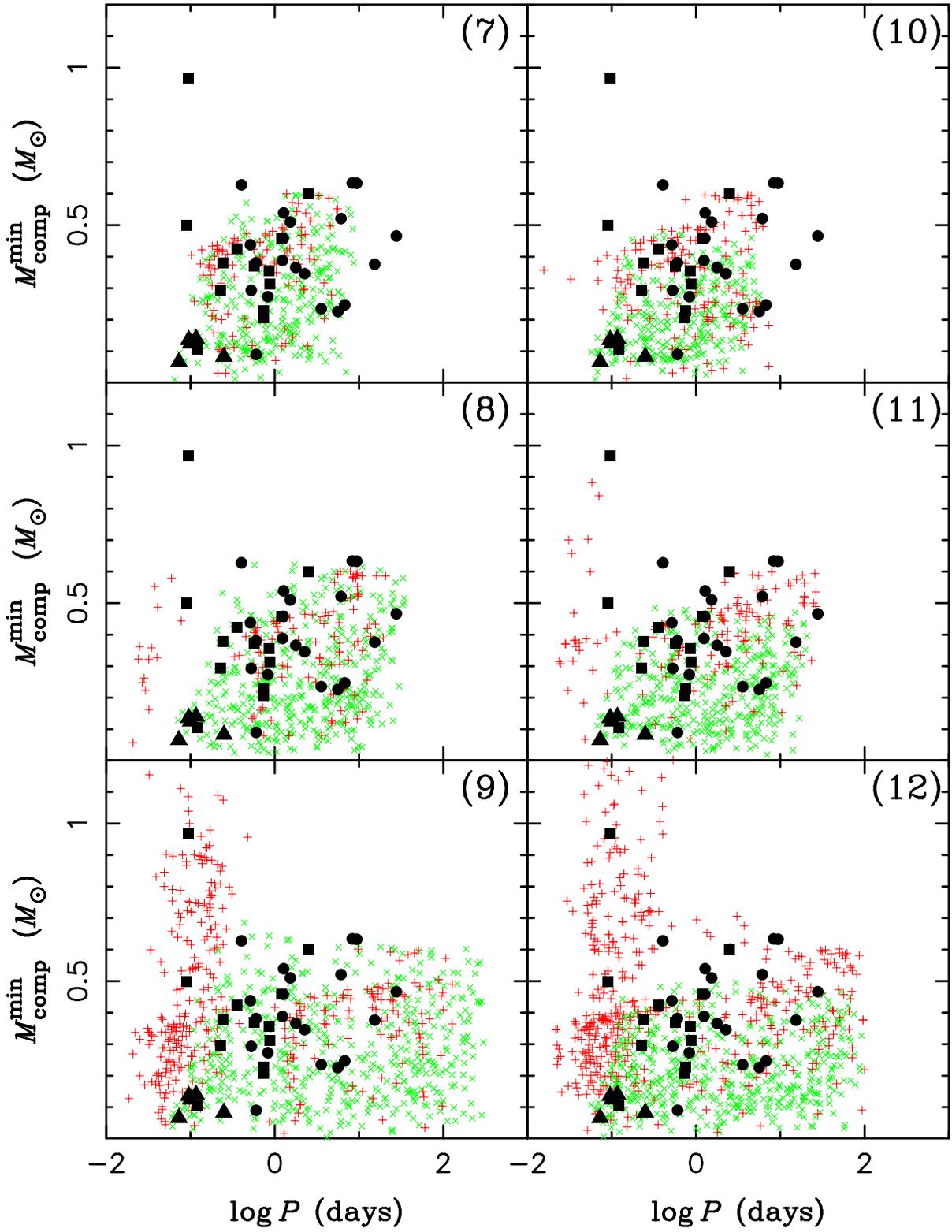}
\caption{
Similar to Figure~\ref{pm1}, but for simulations 7 to 12
as indicated in the upper-right corner of each panel.
}
\label{pm2}
\end{figure*}

The orbital periods $P$ and the mass function, or equivalently the
minimum component masses $M_{\rm comp}^{\rm min}$ (obtained from the
mass function by assuming that the mass of the sdB star is
$0.5\,M_\odot$ and the inclination $\sin i =1$) can be determined
quite precisely from the observational data set of Maxted et al.\ (2001)
and Morales-Rueda et al.\ (2002a,b). Therefore we
choose our best model by mapping the theoretical distribution in the
observational $P$ - $M_{\rm comp}^{\rm min}$ diagram. A
two-dimensional mapping of this type constrains the BPS model much
more severely than any one-dimensional distribution could. We plotted
the $P$ - $M_{\rm comp}$ diagram for all of our simulation sets in
Figures~\ref{pm1} and ~\ref{pm2}.  For a sdB binary produced from our
simulations, we assume that the normal direction of the orbital plane
is randomly distributed in all solid angles.  We then take $\sin i =1$
and $M_{\rm sdB}=0.5\,M_\odot$, no matter what their actually values
are, in order to mimic how this diagram is constructed
observationally.  Here we include the GK selection effect, which is
the major selection effect, in the plotting, but do not consider the
strip selection effect as some of the observational data points are
not affected by it, nor do we include the $K$ selection effect.  In
Figures~\ref{pm1} and ~\ref{pm2}, plus symbols represent sdB binaries
with WD companions and crosses sdB binaries with MS companions (or
red-giant companions) from the simulations.  Filled squares represent
observed sdB binaries with WD companions, filled triangles observed
sdB binaries with MS companions and filled circles observed sdB
binaries of unknown companion type. Visual inspection of these
distributions shows that several of the simulations are in reasonable
agreement with the observational distribution, where simulation set 2
(with $Z=0.02$, a flat mass ratio distribution,
$q_{\rm crit}=1.5$, $\alpha_{\rm CE}=\alpha_{\rm th}=0.75$)
provides the best overall fit.  Based on this comparison, we choose
set 2 as our best-fit model.

Even though the overall distribution in set 2 agrees quite well with
the observed one, the agreement is by no means perfect.  The density
of points near the sdB star KPD 1930+2752 (the filled square at the
top-left corner of panel 2 in Figure~\ref{pm1}) is quite low. This may
be due to the fact that KPD 1930+2752 was specially selected as a
p-mode pulsating star.  It was found by Bill\'eres et al.\
\shortcite{bil00} in a search for pulsators of the EC 14026 variety
\cite{kil97} and may therefore not be a very representative system.
It is also quite possible, perhaps even likely, that the CE ejection
efficiency is not constant for all systems as we assumed.  A higher CE
ejection efficiency for this system could explain it easily (e.g. see
panel 3 of Figure~\ref{pm1}).

Simulation set 2 is quite similar to simulation set 8, except that in
the latter $q_{\rm crit}=1.2$ instead of $q_{\rm crit}=1.5$.  For a
higher $q_{\rm crit}$, the mass donor in the first RLOF phase tends to
be more massive, and therefore the first stable RLOF phase is more
likely to produce WD binaries with high WD masses. The sdB binaries from
the second CE ejection channel therefore tend to have more massive
WD companions (compare panel 2 of Fig.~\ref{pm1} with panel 8 of
Fig.~\ref{pm2}). A value of $q_{\rm crit}$ of 1.5 is higher than the
critical value for dynamical mass transfer obtained from actual binary
evolution calculations (see section 5.1 and Table 3 in Paper I) and may
be in indication for tidally enhanced mass transfer (see the
discussion in section 4.1). It may also by caused by our rather simple
treatment of the first stable RLOF phase on the FGB/AGB. We assume that the
final WD mass is equal to the core mass of the donor at the onset of
the RLOF phase. However, the core mass increases somewhat during
stable RLOF; thus the final WD mass depends on the duration of the
mass-transfer phase which in turn is quite sensitive to $\alpha_{\rm
RLOF}$ \cite{che02}, an effect that needs to be studied further. 


\subsection{sdB binaries with main-sequence companions}

\begin{figure*}
\epsfig{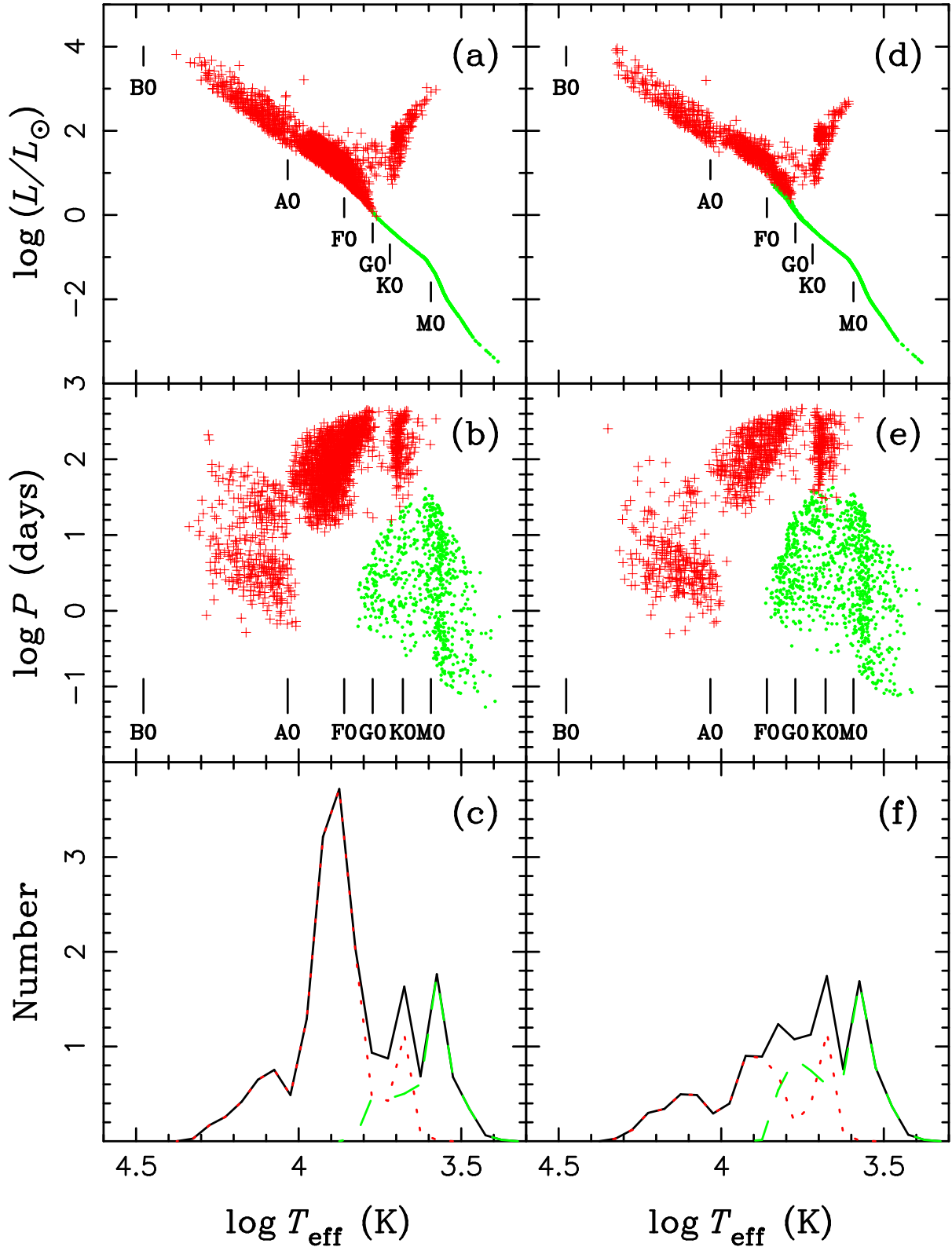}
\caption{The characteristics of the MS secondaries in sdB binaries
from the first CE ejection channel (dots) and the first stable RLOF
channel (crosses) for the best-fit simulations 2 (with $q_{\rm crit} =
1.5$; left panels) and 8 (with $q_{\rm crit} = 1.2$; right panels).
Panels (a) and (d): HRD with spectral types indicated
along the main sequence (based on Zombeck 1990). Panels (b) and (e):
orbital period versus effective temperature. Panels (c)
and (f): the distribution of $\log T_{\rm eff}$; solid curves: both channels;
dashed curves:  the first CE ejection channel only; dotted curves:
the first stable RLOF channel.}
\label{comp}
\end{figure*}

Our best-fit model is mainly constrained by systems that experienced a
CE phase, where in many cases the companion is a white dwarf. Using
our best fits (simulations 2 and 8), our BPS model then makes more
general predictions about the distribution of the properties of sdB
binaries with MS companions.  In Figure~\ref{comp} we present some
of the characteristics of the secondaries in these systems: in the HRD
(panels a and d), in the $P_{\rm orb}$ -- $T_{\rm eff}$ diagram
(panels b and e) and the distributions of $T_{\rm eff}$ (panels c and
f), where the panels on the left represent simulation 2 (with $q_{\rm
crit}=1.5$) and on the right simulation 8 (with $q_{\rm
crit}=1.2$). In the upper panels, dots represents systems from the
first CE ejection channel and plus symbols from the first stable RLOF
channel.

In these panels, one can distinguish four groups of systems, most
clearly seen in the middle panels, corresponding to four peaks in the
$T_{\rm eff}$ distribution in the bottom panels. The systems formed
through the first CE ejection channel (dots) tend to have secondaries
of the latest spectral type (F to M) and have the shortest orbital
periods. Because the secondaries are significantly less massive than
the initial MS mass of the sdB star (due to the $q_{\rm crit}$
constraint), they are essentially unevolved and hence lie close to the
ZAMS in the HRD. Indeed most of the secondaries have spectral type M
(see the ridge in the central panels and the right peak in the bottom
panels). Below an orbital period of $\sim 12\,$hr most and below $\sim
6\,$hr all sdB binaries from the first CE ejection channel have M
dwarf companions, consistent with the fact that all of the 5 sdB
binaries with known MS companions have M type secondaries (see e.g.
Fig.~\ref{bestpm}). The reason for this is simply that these very
low-mass stars have to spiral much deeper into the envelope during the
CE phase before enough orbital energy has been released to eject the
envelope, leading to shorter post-CE orbital periods. Above a period
of $\sim 12\,$hr there is an increasing number of secondaries of
earlier spectral type (as early as F), even though the systems with M
dwarf companions still dominate (this is more prominent in simulation
2 with $q_{\rm crit}=1.5$ than simulation 8 with $q_{\rm crit}=1.2$).

The two groups of systems with the longest orbital periods, mainly
with secondaries of spectral type A to K, are systems from the first
stable RLOF where mass transfer started on the FGB. The gap
between these two groups is just due to the Hertzsprung gap.  The
systems with secondaries of the earliest spectral type (mainly A) are
also systems from the first stable RLOF channel, but where mass
transfer started when the progenitor of the sdB star was in the
Hertzsprung gap. Since in our model the critical mass ratio for stable
RLOF is much larger for systems in the Hertzsprung gap ($q_{\rm crit}
= 3.2$) than for stars on the FGB ($q_{\rm crit} =1.2$/1.5), the
companions in the former can accrete much more mass in the first
stable mass-transfer phase, and hence the secondaries of the sdB stars
tend to be more massive and of earlier spectral type. Furthermore because
of the larger added mass, they are rejuvenated to a larger degree and
therefore are on average somewhat less evolved than secondaries that
accreted from a FGB star (causing the kink around spectral type
A0, most clearly seen in panel d). We note that the precise
distributions of the secondaries in these diagrams are somewhat
sensitive to the assumptions concerning the stable mass-transfer
phase, in particular $\alpha_{\rm RLOF}$, $\alpha$, $q_{\rm crit}$ and
the treatment of the rejuvenation in our simulations. For example, a
lower value of $\alpha_{\rm RLOF}$ would systematically move the
distributions towards lower temperatures.

The comparison of the panels on the left with those on the right
illustrates how dramatically the number of sdB binaries with MS stars
depends on $q_{\rm crit}$. For $q_{\rm crit}=1.5$ (left), these
completely dominate the overall distribution (see the large peak in
panel c), while for $q_{\rm crit}=1.2$ (right) they only form a
significant subset.  This has the interesting implication that
observations of such systems and the determination of their frequency
relative to short-period systems could help constrain $q_{\rm
crit}$, or more generally the conditions for dynamical mass transfer
and/or the importance of tidally enhanced stellar winds.

\subsection{Selection effects}

\begin{figure*}
\epsfig{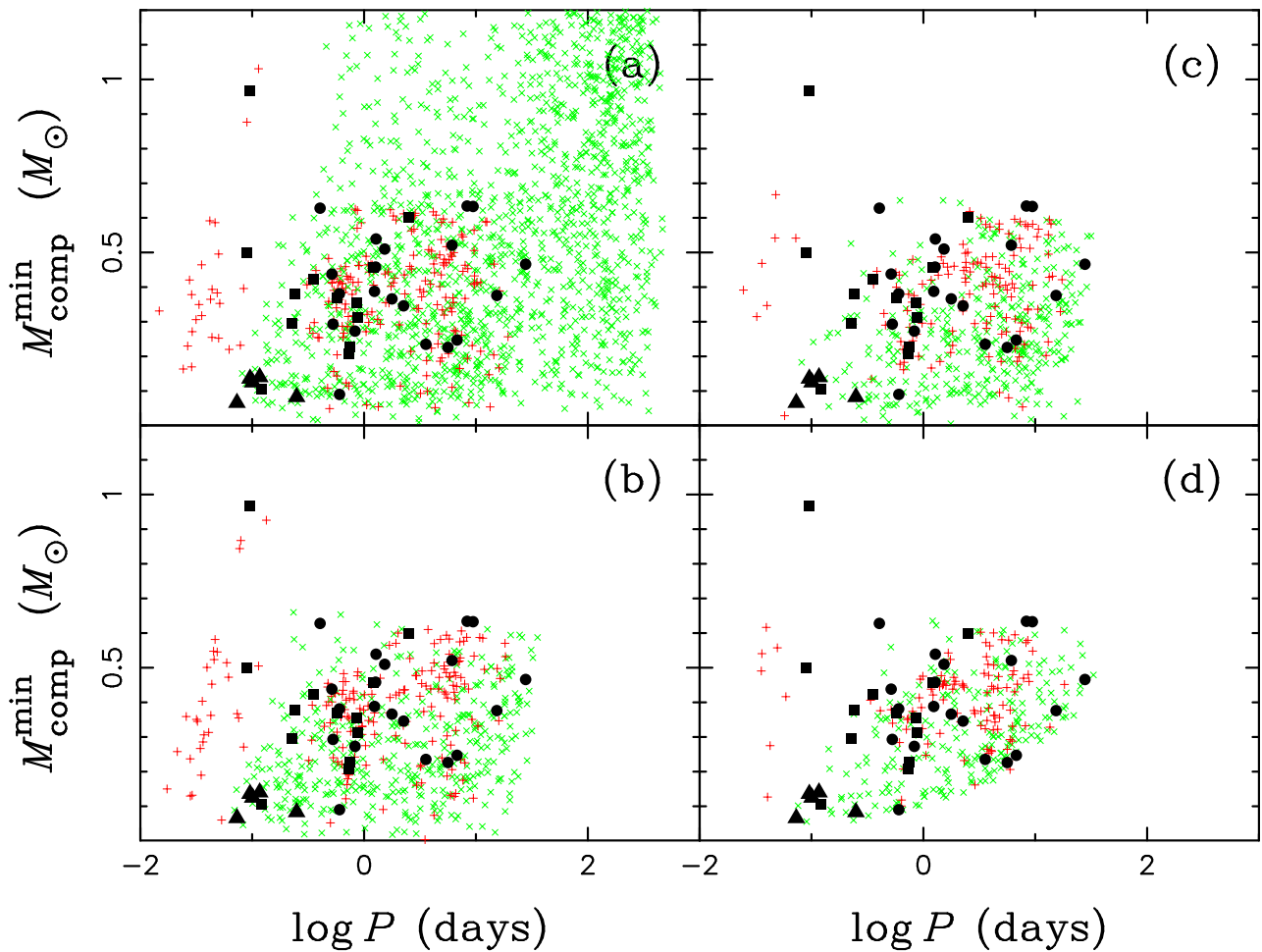}
\caption{
Similar to Figure~\ref{pm1}, but for sdB stars with and without
selection effects from simulation set 2 (the best-fit model): no selection
effects (a), the GK selection effect (b), the GK and the
strip selections effects (c), the GK, the strip and the $K$ selection 
effects (d).
}
\label{bestpm}
\end{figure*}

\begin{figure}
\epsfig{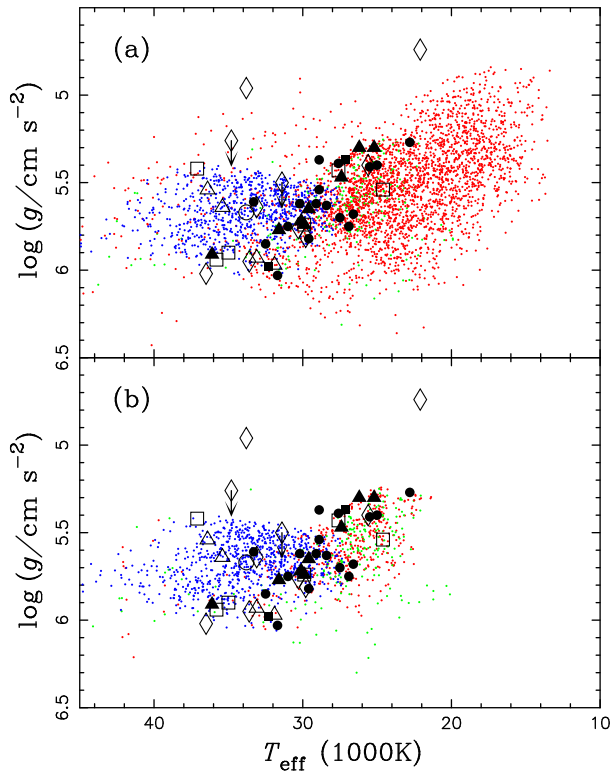}
\caption{
The $T_{\rm eff}$ - $\log g$ diagram for simulation set 2 
(the best-fit model). Dots
represent the results of the simulation.  Filled circles indicate the
position of observed sdB binaries with orbital periods $P_{\rm
orb}<1\,{\rm d}$, solid triangles binaries with $1<P_{\rm
orb}<10\,{\rm d}$, and solid squares binaries with $P_{\rm
orb}>10\,{\rm d}$. Circles show systems that have radial velocity
variations ${\rm d}V>40\,{\rm km}\,{\rm s}^{-1}$, triangles systems
with $20<{\rm d}V<40\,{\rm km}\, {\rm s}^{-1}$, squares systems with
$10<{\rm d}V<20\,{\rm km}\,{\rm s}^{-1}$ and diamonds systems with
${\rm d}V<10\,{\rm km}\,{\rm s}^{-1}$, where ${\rm d}V$ is the maximum
difference between radial velocities measured for a particular object.
Arrows indicate lower limits for $g$.  Panel (a)
does not include selection effects, while in panel (b) the GK
selection effect has been taken into account.
sdB stars from the CE ejection channels are assumed to have envelope
masses between 0.0 and $0.006\,M_\odot$, sdB stars from stable RLOF
channels to have envelope masses between 0.0 and $0.012\,M_\odot$, and
sdB stars from the merger channel to have envelope masses between 0.0 and
$0.002\,M_\odot$. }
\label{besttg}
\end{figure}

\begin{figure}
\epsfig{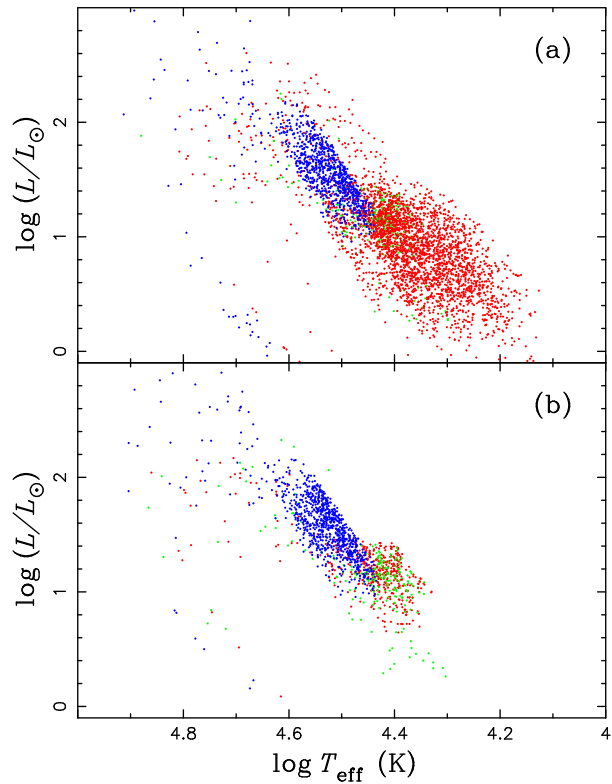}
\caption{
HRD for sdB stars from simulation set 2 (the best-fit model):
panel (a) without any selection effect, and panel (b) with the
GK selection effect. The model assumptions are the same as in 
Figure~\ref{besttg}.
}
\label{besthrd}
\end{figure}

The observed samples of sdB stars are strongly affected by selection
effects. These are relatively well defined for the sample of
Maxted et al.\ \shortcite{max01}. Figure~\ref{bestpm} illustrates
how the selection effects operate for this sample in the $\log
P$ - $M_{\rm comp}^{\rm min}$ diagram and similarly Figures~\ref{besttg} and
\ref{besthrd} show the corresponding effects for the distributions of
sdB stars in the $T_{\rm eff}$ - $\log g$ diagram and the HRD,
respectively (all for simulation 2, our best-fit model).  For
comparison, Figures~\ref{tg-chan} and \ref{hrd-chan} show the
distributions in the $T_{\rm eff}$ - $\log g$ diagram and the HRD for
the different formation channels to illustrate how the selection
effects determine the relative importance of the various channels in
observed samples.

The most important selection effect is the GK selection effect which
we apply in the following way. If a sdB binary has a MS companion and
the effective temperature of the companion is above 4000\,K or the
companion is brighter than the sdB star, the system is excluded. All
sdB binaries from the first stable RLOF are removed in this way in
panel (b) of Figure~\ref{bestpm} as the companions are too massive
(see panel a of Figure~\ref{comp}).  The sdB binaries from the first
CE ejection channel with MS companion masses larger than $0.60\,M_\odot$
for Pop I (or $0.47\,M_\odot$ for a thick-disc population) are also
removed.  The strip effect selects mainly against sdB stars with
masses significantly different from $0.5\,M_\odot$.  This excludes a
significant fraction of sdB stars formed through the merger channel
since these tend to have fairly high masses.  Similarly, sdB stars
which formed from CE ejection channels where their progenitor had a
ZAMS mass larger than the helium flash mass $M_0$ tend to produce sdB
stars with small masses and are also likely to be removed by the strip
selection criterion (see panel c of Fig.~\ref{bestpm}).  Finally,
the $K$ effect selects against sdB stars with long orbital periods and
small companion masses. As all the sdB stars with known orbital
periods have semi-amplitudes $K$ larger than 30\,km\,s$^{-1}$, we
therefore remove sdB binaries with $K$ lower than 30\,km\,s$^{-1} $
(see panel d of Fig.~\ref{bestpm}).

Since the GK effect only excludes sdB binaries, it decreases the
binary fraction of sdB stars (see Table~\ref{percentage}).  On the
other hand the strip selection effect removes both binary sdB stars
and sdB stars from the merger channel. Whether this increases or
decreases the binary fraction relative to the consideration of the GK
effect alone depends sensitively on the CE ejection parameters.  For
simulations with relatively low $\alpha_{\rm CE}$ and $\alpha_{\rm
th}$, the orbital separations of the resulting helium WD pairs are
relatively small leading to a larger merger rate. Since the resulting
single sdB stars tend to be relatively massive, they are mostly
removed by the strip selection effect.

In our simulations, we did not consider any luminosity selection, as
one might expect to be important in a magnitude-limited sample.
Figure~\ref{m-chan} shows the distributions of the masses of sdB stars
from various channels.  The first stable RLOF channel produces a large
fraction of low-mass sdB stars, which tend to have lower luminosity
and should therefore be underrepresented in a magnitude-limited
sample. On the other hand, the merger channel produces a large
fraction of relatively massive sdB stars which one would be able to
detect to larger distances in a magnitude-limited sample.

For reference and comparison we show in Figures~\ref{bestp} to
\ref{semi-k} the distributions of orbital period, the mass of the sdB
star, mass ratio, mass function and radial-velocity semi-amplitude
from all channels for our best-fit model (simulation 2) and show how
these distributions are affected by the selection effects.

\begin{figure*}
\epsfig{file=tg-chan.cps,angle=270,width=17cm}
\caption{
Similar to Figure~\ref{besttg}, but for sdB stars from different
channels in simulation set 2 - the best-fit model 
(no selection effects applied).  Panels
(a), (b), (c) and (d) represent the first CE ejection, the first
stable RLOF, the second CE ejection and the merger channel,
respectively.  }
\label{tg-chan}
\end{figure*}

\begin{figure*}
\epsfig{file=hrd-chan.cps,angle=270,width=17cm}
\caption{
Similar to Figure~\ref{besthrd}, but for sdB stars from different
channels in simulation set 2 - the best-fit model
 (no selection effects applied).  Panels
(a), (b), (c) and (d) represent the first CE ejection, the first
stable RLOF, the second CE ejection and the merger channel,
respectively.  }
\label{hrd-chan}
\end{figure*}

\begin{figure}
\epsfig{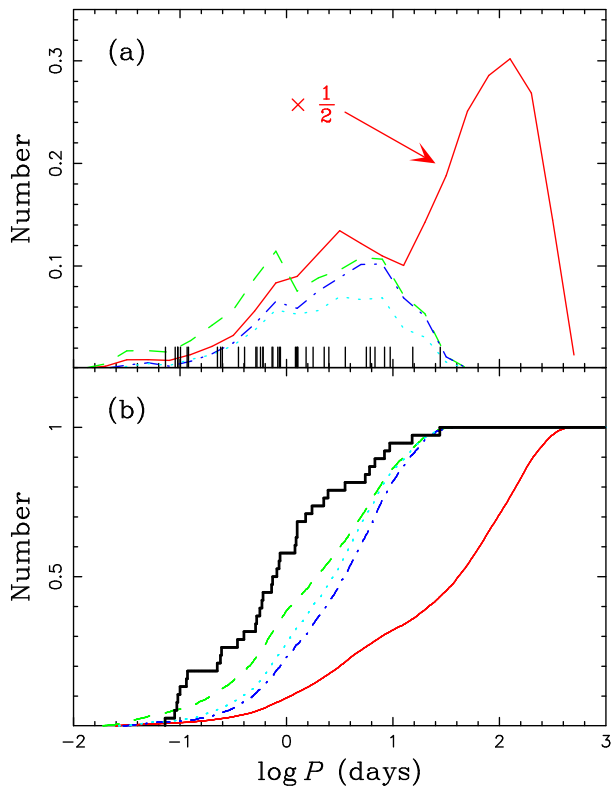}
\caption{
The distribution of orbital periods of sdB stars from all the channels
in simulation set 2 - the best-fit model.  
Panel (a) is a normal distribution while panel
(b) shows a cumulative one.  Solid curves are for sdB stars without
the inclusion of selection effects, while for the dashed curves the GK
selection effect has been applied.  In the dot-dashed curves, both the
GK and the strip selection effects are considered, while for the
dotted curves the GK, the strip and the $K$ selection effects are all
included.  Note that the solid curve in panel (a) has been rescaled by
a factor of 1/2 for clarity Short ticks along the X-axis of panel (a)
indicate the positions of sdB stars in the sample of Maxted et al.\
(2001), while the thick histogram represents the observed distribution
in panel (b).  }
\label{bestp}
\end{figure}
   
\begin{figure}
\epsfig{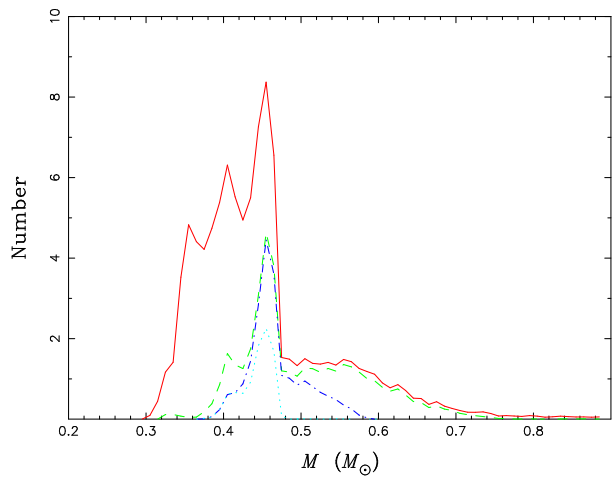}
\caption{
The distribution of
masses of sdB stars from all the channels in simulation set 2 (the best-fit
model);
solid curve: no selection effects; dashed curve: the GK selection effect;
dot-dashed curve: the GK and the strip selection effects; dotted curve:
the GK, the strip and the $K$ selection effects.
}
\label{bestm}
\end{figure}

\begin{figure}
\epsfig{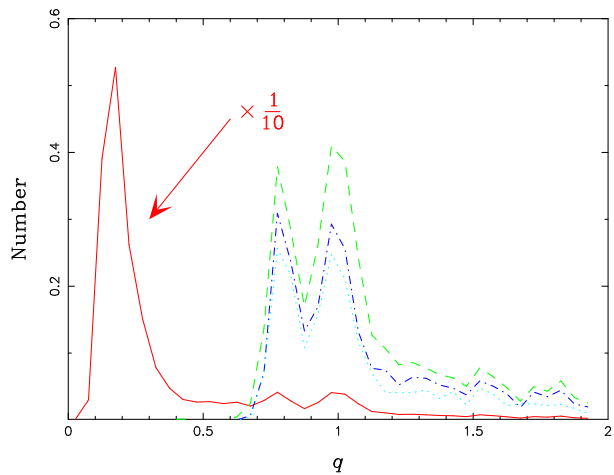}
\caption{
The distribution of mass ratios ($q=M_{\rm sdB}/M_{\rm comp}$) of sdB
binaries from all the binary channels in simulation set 2 (the best-fit
model) when
different selection effects are taken into account: no selection
effects (solid); the GK selection effect (dashed); the GK and the
strip selection effect (dot-dashed); the GK, the strip and the $K$
selection effect (dotted). Note that the solid curve has been rescaled
by a factor of 1/10 for clarity.  }
\label{ratio}
\end{figure}

\begin{figure}
\epsfig{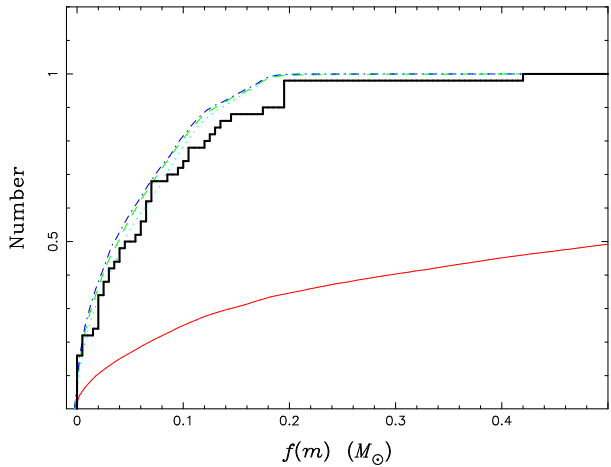}
\caption{
Cumulative distribution of mass functions $f(M)={M_{\rm comp}^3(\sin
i)^3 / (M_{\rm sdB}+M_{\rm comp})^2}$ of sdB stars from all the
channels in simulation set 2 - the best-fit model
 (where we assume that the normal direction of
the orbital plane is uniformly distributed in solid angle).  The
various curves represent the mass function of sdB stars when various
selection effects are taken into account: no selection effects
(solid), the GK selection effect (dashed), the GK and strip selection
effects (dot-dashed), the GK, strip and K selection effects
(dotted). The thick histogram shows the observational distribution of
Maxted et al.\ (2001), Morales-Rueda et al.\ (2002a,b).}
\label{massf}
\end{figure}

\begin{figure}
\epsfig{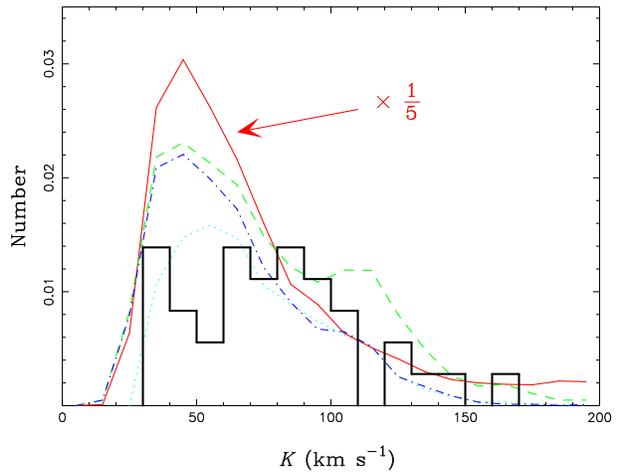}
\caption{
The distribution of the semi-amplitudes $K$ of radial velocities for
sdB binaries from all the binary channels in simulation set 2 - the best-fit
model (where
we assume that the normal direction of the orbital plane is uniformly
distributed in solid angle).  The solid curve represents sdB stars
without any selection effect considered (the line has been rescaled by
a factor of 1/5 for clarity), the dashed curve includes the GK
selection effect, the dot-dashed curve both the GK and the strip
selection effects, and the dotted curve the GK, the strip and the $K$
selection effects. The thick histogram represents the observational
sample of Maxted et al.\ 2001, Morales-Rueda et al.\ 2002a,b).  }
\label{semi-k}
\end{figure}

\subsection{Expected birthrates and total numbers}

As Table~1 shows, the predicted birthrate of Pop I sdB stars from all
channels is in the range $0.014$\,--\,$0.063\,{\rm yr}^{-1}$ for the
whole Galaxy, where our best models (simulations 2 and 8) give a rate
of $\sim 0.05\, {\rm yr}^{-1}$. The formation rate from the merger
channel alone, which produces single sdB stars, is in the range of
0.003\,--\,$0.017\,{\rm yr}^{-1}$, somewhat lower than the estimate of
Tutukov \& Yungelson (1990) who obtained a rate of $0.029\,{\rm
yr}^{-1}$. By taking an effective Galactic volume of $5\times
10^{11}\,{\rm pc}^3$ \cite{zom90}, this can be converted into an
average birthrate per pc$^3$ of $2.8$\,--\,$12.6\times 10^{-14}\,{\rm
pc}^{-3}\,{\rm yr}^{-1}$ or $10\times 10^{-14}\,{\rm pc}^{-3}\,{\rm
yr}^{-1}$ for the best model.  When convolved with the lifetime of the
sdB phase, these rates imply a total number of sdB stars in the Galaxy
of $2.4\times 10^6$\,--\,$9\times 10^6$, or a space number density of
0.5\,--\,$1.9\times 10^{-5}\,{\rm pc}^{-3}$, where our best estimates
are $\sim 6\times 10^6$ and $\sim 1\times 10^{-5}\,{\rm pc}^{-3}$,
respectively.  Including the GK selection effect reduces both the
number and the density of selected sdB stars by about a factor of
2. The inclusion of the strip selection effect further halves these
numbers. For a thick-disc population, the birthrate and the total
number of sdB stars would be higher than for Pop I if we adopted similar
model parameters for both populations (in particular, using the same
star-formation rate).

\subsection{Comparison with observations}

\begin{figure}
\epsfig{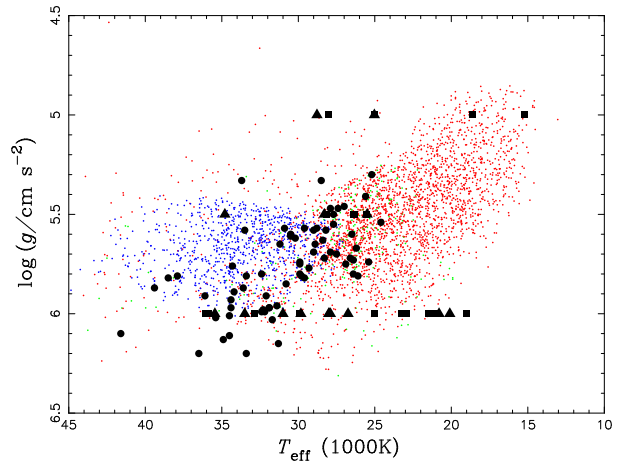}
\caption{The
 $T_{\rm eff}$ - $\log g$ diagram for simulation set 2 - the best-fit model. 
 Filled
 circles show the position of observed sdB stars from Saffer et al.\
 (1994), filled triangles and filled squares represent single and binary
 sdB stars, respectively, from the observations of Aznar Cuadrado \&
 Jeffery (2001).  }
\label{besttgaj}
\end{figure}

\begin{figure}
\epsfig{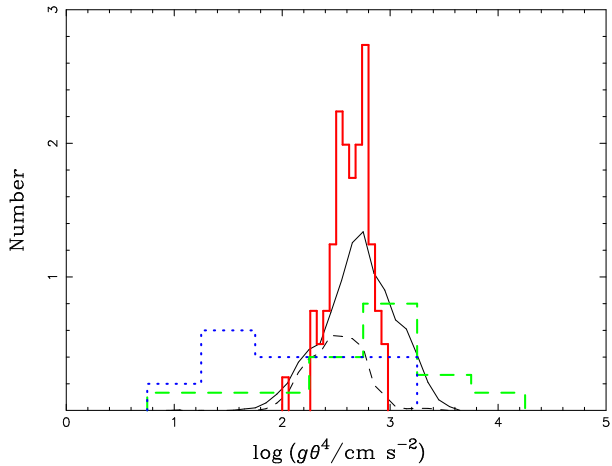}
\caption{
The distribution of $\log (g\theta^4)$, where $g$ is the surface
gravity of a sdB star, and $\theta=5040\,{\rm K}/T_{\rm eff}$.  The
solid histogram is obtained from the 68 sdB stars observed by Saffer
et al.\ (1994), while the dashed and dotted histograms are based on
the observations of 15 sdB stars and 10 sdO stars, respectively, by
Ulla \& Thejll (1998). The solid curve represents the sdB stars from
simulation set 2 (the best-fit model) 
without consideration of selection effects, while the
dashed curve includes the GK selection effect.}
\label{theta}
\end{figure}

Heber \shortcite{heb86} estimated the birthrate of sdB stars to be
$\sim 2\times 10^{-14}\,{\rm pc}^{-3} \,{\rm yr}^{-1}$ and the space
density to be $\sim 4\times 10^{-6} \,{\rm pc}^{-3}$.  Downes
\shortcite{dow86} derived a space density of $\sim 2\times
10^{-6}\,{\rm pc}^{-3}$ from observations, while Bixler, Bowyer \&
Laget \shortcite{bix91} obtained a density of $\sim 3.3\times
10^{-6}\,{\rm pc}^{-3}$. Most other studies gave similar values,
although Villeneuve et al.\ \shortcite{vil95}, using a much larger
scale height, obtained a birthrate density and a space density which
are by a factor of $\sim 5$--\,10 lower than the previous estimates,
although this would not affect the overall birthrate estimate. The
observational estimates are in reasonable agreement with our
theoretical estimates, in particular after selection effects have been
taken into account.

Observationally, over half of the sdB stars are in binaries with
MS/giant companions \cite{all94,azn01}. The sdB stars with
MS/giant companions constitute 63\,--\,88 per cent in our simulations.
More than two thirds of the sdB candidates of Maxted et al.\
\shortcite{max01} are binaries with short orbital periods. With
the GK effect, the sdB binaries with short orbital periods produced
from our simulations constitute 41\,--\,86 per cent of the observable
population.  In the observational data set of Maxted et al.\
\shortcite{max01}, 13 of 18 (or 72 per cent) sdB binaries with known
companion types have WD companions \cite{mor02b}, although the
majority of sdB stars in the sample are presently of unknown type. The
relative number of systems with WD and MS secondaries will allow to
further refine the BPS model, in particular $q_{\rm crit}$; however,
as we have shown, these numbers are strongly affected by the selection
effects.

Figure~\ref{besttg} displays a comparison of our best model (simulation 2)
with the observations of Maxted et al.\ \shortcite{max01} in the $T_{\rm eff}$
-- $\log g$ diagram. The distribution of observed systems (as indicated 
by large symbols) matches the simulated one  (as indicated by dots) 
quite well after the GK selection effect has been taken into account.  PG
1051+501 and PG 1553+273 (the two top diamonds) may originate from the
first stable RLOF channel and may have relatively large hydrogen-rich
envelopes (see also panel b of Figure~\ref{tg-chan}).
In the simulation, sdB stars from the CE ejection channels are 
assumed to have envelope masses between 0.0 and $0.006\,M_\odot$, 
sdB stars from stable RLOF channels to have envelope masses 
between 0.0 and $0.012\,M_\odot$, and sdB stars from the merger 
channel to have envelope masses between 0.0 and $0.002\,M_\odot$.
Brown et al.\ \shortcite{bro01} pointed out that
the envelope composition can be changed 
dramatically (e.g. from hydrogen-rich to helium-rich) 
due to helium-flash-induced mixing between the interior and the envelope. 
The hydrogen mixed into the hot He-burning interior is burned rapidly and
the mass of the hydrogen-rich envelope can be reduced.
Such mixing is found for sdB stars evolving to steady core helium burning 
with hydrogen-rich envelopes of masses
$\sim 0.0006M_\odot$. 
The mixing may also occur in a binary model and have a small effect
on envelope masses, while the assumption on envelope masses is rather
ad hoc in the simulation.
Figure~\ref{besttgaj} compares the observations of Saffer et al.\
\shortcite{saf94} and Aznar Cuadrado \& Jeffery \shortcite{azn01} with
the sdB stars from our best model (as indicated by dots).  Filled
circles show the position of observed sdB stars from Saffer et al.\
\shortcite{saf94}, filled triangles and filled squares represent
single and binary sdB stars, respectively, from the observation of
Aznar Cuadrado \& Jeffery \shortcite{azn01}.  The candidates of Saffer
et al.\ \shortcite{saf94} were taken from the PG catalogue and therefore
suffer from the GK selection effect, while the candidates of Aznar
Cuadrado \& Jeffery \shortcite{azn01} were taken from the IUE archive
and are affected by uncertain selection effects. Aznar Cuadrado \&
Jeffery \shortcite{azn01} used a grid of high-gravity helium-deficient
model atmospheres \cite{odo97} to determine the atmospheric
parameters. Their grid has a spacing of 2000\,K in $T_{\rm eff}$ and a
spacing of 0.5 dex in $\log g$.  Therefore, the $\log g$ values of
their measured sdB stars have only three discreet values: 5.0, 5.5 and
6.0. With these limitations in mind, we conclude that our
simulations can reasonably explain their observations.

The distribution of orbital periods of sdB binaries \cite{max01} is
explained reasonably well in Figure~\ref{bestp} after the various
selection effects have been applied. However, the distribution also
suggests that the observed sample may be missing some of the sdB
binaries with relatively long orbital periods.

The distribution of the masses of sdB stars from simulation set 2 
(the best-fit model) is
plotted in Figure~\ref{bestm}; note in particular how narrow the mass
distribution becomes once selection effects have been taken into
account. This also implies that real intrinsic mass distribution of
sdB stars should be much wider than the observed one. Since it is
difficult to measure the mass directly from observation, we also
plot in Figure~\ref{theta} 
the distribution of $\log (g\theta^4)$ (where $\theta=5040{\rm
K}/T_{\rm eff}$) for the sdB stars from simulation set 2 and
histograms for 68 sdB stars observed by Saffer et al.\
\shortcite{saf94}, 15 sdB stars and 10 sdO stars observed by Ulla \&
Thejll \shortcite{ull98}. The quantity $g\theta^4$ is approximately
constant for sdB stars of a given mass (since it is proportional to
the mass -- luminosity ratio; Greenstein \& Sargent 1974, also see
Fig.~3 of Paper I), and therefore the distribution of this quantity
provides some information on the mass distribution. The distribution
from simulation set 2 is consistent with that from Saffer et al.\
\shortcite{saf94} after inclusion of the GK selection effect and
taking into account the fact that some of the sdB stars from
simulation set 2 are actually sdO stars. Those sdO stars come from the
merger channel and are massive and therefore have low values of
$g\theta^4$.

The mass-ratio distribution (see Figure~\ref{ratio}) has three peaks:
the first peak is caused by sdB star from the first stable RLOF
channel, the second and the third are due to the bimodal distribution
of WD masses in the second CE ejection channel. However, it is the
mass function rather than the mass ratio that can be measured directly
from observations.  The observed mass function distribution (see
Figure~\ref{massf}) is well explained after application of the GK
selection effect.

Figure~\ref{semi-k} gives the distributions of radial-velocity
semi-amplitudes $K$ of sdB binaries.  The comparison between the
theoretical and the observed distributions suggests that some sdB
stars with a low value of $K$ are missing in the observed samples.

\subsection{Comparison with the results of previous studies}

D'Cruz et al.\ \shortcite{dcr96} tried to understand the 
formation of sdB stars by employing and varying the Reimers mass-loss
formula near the tip of the FGB. In their picture, it was a stellar
wind that peeled off the hydrogen envelope of a FGB star before helium
ignition, which then occurred at much higher $T_{\rm eff}$ leading to
the formation of a sdB star.  They assumed a broad distribution in the
Reimers coefficient $\eta$ in order to explain the observations. The
value of $\eta$ has to be 2\,--\,3 times larger in some stars to produce
sdB stars than to produce normal horizontal-branch stars. At present,
there is no theoretical justification for such a range of $\eta$
values for single stars. On the other hand, our model provides a
natural way to produce sdB stars without tuning the Reimers
coefficient. Binary interactions naturally expose the
hydrogen-exhausted cores of FGB stars either by stable RLOF or CE
ejection. An enhanced wind, as required by D'Cruz et al.\
\shortcite{dcr96}, may be possible in binary systems since the stellar
wind may be tidally enhanced due to the proximity of a companion star
\cite{egg89b,han95b}. sdB stars produced in this way would be  binaries with
relatively long orbital periods. We have not included a tidally
enhanced stellar wind since we did not want to introduce further
uncertainties into the modelling. Nevertheless, this channel certainly
needs to be studied further even though the channel may ultimately
turn out not to be very significant since it probably requires
significant fine-tuning of the stellar wind parameters (i.e. it
requires a fairly narrow range of binary separations).

Webbink \shortcite{web84}, Iben \& Tutukov \shortcite{ibe86}, Tutukov
\& Yungelson (1990) and Iben et al.\ (1997)
have investigated in detail the merger channel for the formation of
sdB stars.  Their most recent estimate (Iben et al.\ 1997) for the
merger rate of helium WDs in the Galaxy is $\sim 0.02\,{\rm yr}^{-1}$,
which is quite similar to our estimate of 0.003\,--\,$0.017\,{\rm
yr}^{-1}$. Iben et al.\ \shortcite{ibe97} did not specifically examine
whether the merger product would ignite helium and hence become a sdB
star. Nevertheless, their results are consistent with ours, since we
found in our simulations that in fact most merger products of helium
WD pairs ignite helium.  Han
\shortcite{han98} gave a birthrate of 0.002\,--\,$0.014\,{\rm
yr}^{-1}$ in his study on the formation of double degenerates. Our new
birthrate is slightly larger than his, mainly because we adopted a
higher value for $q_{\rm crit}$ for the first stable RLOF phase, which
makes the second CE ejection channel more likely and ultimately
increases the merger rate. The distribution of masses of helium WD
mergers in the present paper (see Fig.~\ref{m-chan}) is similar to the
distribution in Figure~6 of Han (1998).

Mengel, Norris \& Gross \shortcite{men76} have modelled the
conservative evolution of a binary system with initial masses of
$0.80\,M_\odot$ and $0.78\,M_\odot$ for the primary and the secondary,
respectively, for a composition with $X=0.73$ and $Z=0.001$. They also
found that there is a range of initial separations for which a sdB
star is formed as a result of stable and conservative RLOF.  The sdB
star formed in their calculations had a mass $\sim 0.5\,M_{\odot}$ and an
orbital period of $\sim 300\,{\rm d}$. Both the mass and the period
fall inside the range of sdB stars from the first stable RLOF channel
in our simulations although we use higher metallicities and assume
non-conservative mass transfer.

\subsection{Further observational tests to the model}

In this paper we used the well-defined sample of sdB stars of Maxted
et al.\ \shortcite{max01} and Morales-Rueda et al.\ (2002a,b) to calibrate
our BPS model. However, because of the design of the sample and
various selection effects, it only comprises a subset of the whole
population of sdB stars included in the BPS model. Hence our model can
be used to make predictions about the wider population of sdB
stars. Extending the observational sample should then allow to test
these predictions and to help refine some of the BPS parameters that
are presently not well constrained.

As one can see from Figures~\ref{besttg} and \ref{tg-chan}, there are
a few observed sdB stars (in fact sdO stars) with $\log g\sim 5.75$
and $T_{\rm eff}\sim 40000\,{\rm K}$. Their position in the $T_{\rm
eff}$ -- $\log g$ suggests that they are more massive than $\sim
0.5\,M_{\odot}$ (see Fig.~2d of Paper I).  Since the GK selection
effect tends to eliminate sdB stars from the first stable RLOF
channel, most of these are likely to be single sdB stars formed
from the merger channel (although one of them, PG0839+399, is a binary
with an orbital period of 5.622\,d).

Our model predicts that a large fraction, perhaps the majority of the
intrinsic population of sdB stars have MS companions. In particular,
Figure~\ref{comp} shows that sdB stars from the first stable RLOF can
have companions with a spectral type as early as B. The predicted
numbers of sdB stars with B, A or F type companions in the Galaxy at
the current epoch are 0.69, 2.4, 0.89 million, respectively, for
simulation set 2 (our best-fit model with $q_{\rm crit}=1.5$) 
or 0.50, 0.61, 0.56 
million for simulation set 8 (with $q_{\rm crit}=1.2$).
The numbers of B, A, or F type stars in the Galaxy at the current
epoch are 34, 314, 1898 million, respectively, in our BPS model. These
numbers imply that 2.0 per cent of B type stars, 0.75 per cent of A
type stars and 0.047 per cent of F type stars should have sdB
companions, respectively, for simulation set 2. For simulation set 8,
the corresponding numbers are 1.4 per cent for B type stars, 0.19 per
cent for A type stars and 0.030 per cent for F type stars. The main
difference between simulation sets 2 and 8 is the value of $q_{\rm
crit}$ (1.5 in set 2 and 1.2 in set 8).  These percentages demonstrate
(also see the discussion in section~7.5) that the percentage of A type
stars with sdB companions is quite sensitive to $q_{\rm crit}$, the
critical mass ratio above which mass transfer is dynamically unstable
on the FGB or AGB (The sensitivity is due to the fact that
such systems have experienced stable RLOF on the FGB).  
Observations of A type stars with sdB companions
may therefore help to constrain $q_{\rm crit}$, a basic and important
parameter in any BPS model.

\section{Conclusion}

In this paper we have presented a comprehensive BPS study for the
formation of sdB stars and investigated the importance of the various
evolutionary channels that lead to the formation of sdB stars. We
studied the roles of both the theoretical model parameters and the
observational selection effects.  We obtained birthrates for sdB stars
of 0.014 -- 0.063\,$\,{\rm yr}^{-1}$, a total number of sdB stars in the
Galaxy of 2.4 -- 9.5 million and a sdB binary fraction of 76 -- 89
per cent.  The distribution of orbital periods ranges from 0.5\ hr to
500\ d, possibly with three peaks at $\sim 2.4$\,hr, $\sim 2$\,d and
$\sim 100$\,d.  The distribution of masses has a fairly
wide range from $0.3\,M_\odot$ to $0.8\,M_\odot$ with a major peak
near $0.46\,M_\odot$.

Comparing our simulations to observed samples of sdB stars, we found a
best-fit model that explains the observed distribution quite
satisfactorily with very reasonable theoretical parameters. Based on
this best-fit model, we conclude:
\begin{enumerate}
\item
The first RLOF needs to be more stable for a wider range of parameters
than is commonly assumed, either because of a higher critical mass
ratio $q_{\rm crit}$ for the occurrence of dynamical mass transfer or
a tidally enhanced stellar wind. This suggests that the criterion for
stable RLOF needs to be studied further.
\item
The first stable RLOF is non-conservative, and the mass lost from the
system carries away a specific angular momentum similar to that of the
system.
\item 
In agreement with earlier studies, we find that the common-envelope
ejection is a very efficient process, though the values of
$\alpha_{\rm CE}$ and $\alpha_{\rm th}$ cannot yet be precisely
determined\footnote{Soker \& Harpaz (2002) have recently criticized our
interpretation of this result in Paper I, specifically that this
implies that a significant fraction of the thermal energy (including
the ionization energy) can be used in the CE ejection. Their arguments
are based on a very different view of the physical processes involved
in the CE ejection process -- in our view a quite inappropriate
one. We will address the relevant issues in the context of a more
general theoretical paper on the physics of CE ejection in a future
publication. Here we just note that the approach taken in this paper,
as well as in Paper I, is entirely empirical and is independent
of theoretical pre-conceptions.}.
\item
Our best model explains the observed properties of sdB stars
quite satisfactorily (in particular, the
$\log P$ -- $M_{\rm comp}$ diagram, the $T_{\rm eff}$ -- $\log g$
diagram, the orbital period distribution, the $\log (g\theta^4)$ distribution,
the mass function distribution, the  binary fraction of sdB stars, the birth
rates, the space number densities etc.).
\item
Our best-fit model predicts a much wider distribution of masses for
sdB stars than is commonly assumed.  It also predicts that some B, A
and F type stars have sdB companions and that the percentage of A type
stars with sdB companions can be used to constrain the critical mass
ratio for stable RLOF on the FGB.
\end{enumerate}

\section*{Acknowledgements}
We thank Dr.\ A.\ Lynas-Gray for helpful discussions. 
We are grateful to an anonymous referee for his/her useful comments.
This work was in part supported by a Royal Society UK-China Joint Project
Grant (Ph.P and Z.H.), the Chinese National Science Foundation under 
Grant No.\ 19925312, 10073009 and NKBRSF No. 19990754 (Z.H.).

\end{document}